\documentclass[twocolumn]{aastex62}
\usepackage{appendix}
\usepackage{natbib}
\usepackage{amsmath}
\usepackage{booktabs}
\usepackage{multirow}
\usepackage{color,subfigure,lineno}
\newcommand{\RNum}[1]{\uppercase\expandafter{\romannumeral #1\relax}}
\usepackage{comment}

\newcommand{\hbeta}{H{$\beta$}}
\newcommand{\halpha}{H{$\alpha$}}

\def \OIII {[O\,{\sc iii}]}

\newcommand{\comments}[1]{}

\begin{document}

\title{\bf{Seoul National University AGN Monitoring Project. V. Velocity-resolved \hbeta\ Reverberation Mapping and Evidence of Kinematics Evolution} }

\author[0000-0002-2052-6400]{Shu Wang}
\affiliation{Astronomy Program, Department of Physics and Astronomy, Seoul National University, Seoul, 08826, Republic of Korea; jhwoo@snu.ac.kr}

\author[0000-0002-8055-5465]{Jong-Hak Woo}
\affiliation{Astronomy Program, Department of Physics and Astronomy, Seoul National University, Seoul, 08826, Republic of Korea; jhwoo@snu.ac.kr}

\author[0000-0002-3026-0562]{Aaron J.\ Barth}
\affil{Department of Physics and Astronomy, 4129 Frederick Reines Hall, University of California, Irvine, CA, 92697-4575, USA}

\author[0000-0003-2064-0518]{Vardha N. Bennert}
\affil{Physics Department, California Polytechnic State University, San Luis Obispo, CA 93407, USA}

\author[0000-0001-5802-6041]{Elena Gallo} 
\affil{Department of Astronomy, University of Michigan, Ann Arbor, MI 48109, USA}

\author[0000-0002-2397-206X]{Edmund Hodges-Kluck} 
\affil{NASA/GSFC, Code 662, Greenbelt, MD 20771, USA}

\author[0000-0002-3560-0781]{Minjin Kim}
\affil{Major in Astronomy and Atmospheric Sciences, Kyungpook National University, Daegu 41566, Republic of Korea}

\author[0000-0002-8377-9667]{Suvendu Rakshit}
\affiliation{Aryabhatta Research Institute of Observational Sciences, Manora Peak, Nainital-263001, Uttarakhand, India}

\author[0000-0002-8460-0390]{Tommaso Treu}
\affil{Department of Physics and Astronomy, University of California, Los Angeles, CA 90095-1547, USA}

\author[0000-0003-2010-8521]{Hojin Cho}
\affiliation{Astronomy Program, Department of Physics and Astronomy, Seoul National University, Seoul, 08826, Republic of Korea; jhwoo@snu.ac.kr}
\affiliation{Department of Physics \& Astronomy, Texas Tech University, Lubbock TX, 79409-1051, USA}

\author[0000-0003-2632-8875]{Kyle M. Kabasares}
\affiliation{Bay Area Environmental Research Institute, Ames Research Center, Moffett Field, CA 94035, USA}
\affiliation{Ames Research Center, National Aeronautics and Space Administration, Moffett Field, CA 94035, USA}
\affiliation{Department of Physics and Astronomy, 4129 Frederick Reines Hall, University of California, Irvine, CA, 92697-4575, USA}

\author[0000-0001-6919-1237]{Matthew A. Malkan} 
\affil{Department of Physics and Astronomy, University of California, Los Angeles, CA 90095-1547, USA}

\author[0000-0001-9957-6349]{Amit Kumar Mandal}
\affiliation{Astronomy Program, Department of Physics and Astronomy, Seoul National University, Seoul, 08826, Republic of Korea; jhwoo@snu.ac.kr}

\author[0000-0002-4704-3230]{Donghoon Son}
\affiliation{Astronomy Program, Department of Physics and Astronomy, Seoul National University, Seoul, 08826, Republic of Korea; jhwoo@snu.ac.kr}

\author[0000-0002-1912-0024]{Vivian U}
\affil{Department of Physics and Astronomy, 4129 Frederick Reines Hall, University of California, Irvine, CA, 92697-4575, USA}

\author[0000-0002-1961-6361]{Lizvette Villafana}
\affil{Department of Physics and Astronomy, University of California, Los Angeles, CA 90095-1547, USA}
\affil{Physics Department, California Polytechnic State University, San Luis Obispo, CA 93407, USA}


\begin{abstract}
We present velocity-resolved  reverberation lags of \hbeta\ for 20 active galactic nuclei (AGNs) from the Seoul National University AGN Monitoring Project. We detect unambiguous velocity-resolved structures in 12 AGNs, among which eight objects exhibit symmetric structures, two objects show inflow-like characteristics, and two objects display outflow-like signatures. For two AGNs, we successfully measure the velocity-resolved lags in different years, revealing  evidence of evolving broad-line region (BLR) kinematics.  By combining  our sample with the literature velocity-resolved lags, we find that the symmetric velocity-resolved lags are the most common ($40$\%) type among this sample. The frequency of inflow kinematics is also notable ($20$\%), while outflow kinematics are less common ($11$\%).
Our sample significantly expands the previous velocity-resolved reverberation mapping sample in the high-luminosity regime, enabling us to constrain BLR kinematics across a large dynamic range of luminosity.

\end{abstract}


\section{Introduction}

Reverberation mapping (RM) has emerged as a powerful tool for probing the radius and structure of broad-line regions (BLRs) in active galactic nuclei \citep[AGNs;][]{Blandford_McKee_1982, Peterson93}.  This technique enables the measurement of the time delay between the variation of continuum and the response of the broad emission line, which is used to trace the average BLR radius ($R_{\rm BLR}$). By combining the velocity indicated by the broad-line widths ($\Delta V$), the black hole (BH) mass can be estimated under the virial assumption, i.e.,  $M_{\rm BH}=f\; \frac{R_{\rm BLR} \Delta V^2}{G}$, where $f$ is the virial coefficient that accounts for the BLR's inclination, geometry and kinematics.  More than 200 virial BH masses have been obtained based on \hbeta\ RM over the past 30 yr
\citep[e.g.,][]{Kaspi00, Peterson2004, Bentz09b,  Denney09b, Denney10, Barth11, Wang14, Du14, Du16b, Du18,  Grier17b, Fausnaugh17, DeRosa18, U22, Marlik23, Shen23Arxiv, Woo24, Li-Y24}. RM studies further reveal a scaling relationship between the BLR size and AGN continuum luminosity \citep[e.g.,][]{Kaspi00, Bentz09,Bentz13}, enabling the single-epoch (SE) BH mass estimation that can be easily applied to large samples and high redshifts.

Despite the significant advance in BH mass estimations, our understanding of the BLR geometry and kinematics is still limited.  The virial assumption, which is the foundation of RM and SE BH mass estimation, has been  primarily examined in local Seyferts \citep[e.g.,][]{Peterson99,Peterson00,Lu16}. Further investigation is needed for AGNs spanning a wider range of luminosities and redshifts.  
One effective way to address this issue is studying the BLR transfer function \citep[or velocity-delayed map; e.g.,][]{Robinson90,Welsh91,Perez92}, which describes how BLR clouds respond to the continuum variations in a  two-dimensional velocity--delay space.
Several inversion techniques have been developed to recover the BLR transfer function, including maximum entropy methods \citep{Horne91,Horne94} and regularized linear inversion  \citep{Krolik_Done95,Skielboe15}.  Alternatively, forward modeling of BLRs can be employed to search for the best phenomenological model that describes the observations.  This method enables independent constraints on $M_{\rm BH}$ and the virial coefficient for individual AGNs  \citep[e.g.,][]{Pancoast11,Pancoast14b,Pancoast18,Li13,Grier17a,Williams18,Bentz21,Bentz22,Bentz23b,Villafana22,Villafana23,Stone24}. 
However, both approaches require high-fidelity datasets, e.g., high cadence, long duration and high signal-to-noise ratio spectra. As a result, the number of RM AGNs with such measurements is still very limited.

A straightforward and efficient approach is to divide the broad emission line into different velocity bins and measure the lags for each bin, a method known as velocity-resolved lags. The structure of these velocity-resolved lags offers a qualitative description of the BLR kinematics \citep[e.g.,][]{Gaskell88,Crenshaw90,Bentz09b,Denney09b,Denney10}. 
For instance, a symmetric structure with the line center lagging behind the line wing, is consistent with disk-like rotation.
A structure where blue velocity bins show longer lags than red velocity bins suggests inflow, while the opposite pattern—red velocity bins showing longter lags than blue velocity bins—indicates outflow.

Over recent years, velocity-resolved lags have been successfully measured for approximately 70 AGNs \citep{Kollatschny03, Bentz09b, Denney09b, Denney10, Barth11, Doroshenko12, Grier13, Pei14, Valenti15, Du16, Lu16, Park17, Pei17, Du18, DeRosa18, Zhang19, Brotherton20, Hlabathe20, Hu20b, Bentz21,Bentz23a, Bentz23b, Lu21, Feng21a, Feng21b, Feng24, Li-SS22, Li-S24, Bao22,Chen23,Zastrocky24,Li-Y24}, which significantly enhances our knowledge about the BLR kinematics. While the most commonly observed type is disk-like rotation, these studies reveal that AGN BLRs do exhibit diverse kinematics \citep[e.g.,][]{Du16, U22}. No significant luminosity-dependent trend of inferred BLR kinematics is found \citep{U22}, based on the current velocity-resolved RM sample which consists of mostly low-to-moderate AGNs (e.g., $L_{5100}\leq 10^{44}$ erg s$^{-1}$). It requires the inclusion of higher luminosity AGNs to provide direct examinations for BH mass estimation of high-redshift AGNs.

In the velocity-resolved RM sample,  some AGNs have been observed multiple times, enabling the study of the time evolution of BLR kinematics.  \citet{Pancoast18} finds that the BLR kinematics of Arp 151  is stable over 7 yr. Conversely,   \citet{DeRosa18} finds significant difference when comparing the velocity-resolved structure from 2012 observation with that from the 2007 observation \citep{Denney09b} for two AGNs, i.e., NGC 3516 and NGC 3227. NGC 4151 also displays back-and-forth variations among  disk-like rotation, inflow, and outflow kinematics over the past two decades \citep{Chen23}. In addition, the Monitoring AGNs with \hbeta\ Asymmetry (MAHA) project \citep{Du18b,Brotherton20,Bao22,Zastrocky24} has observed evidence of evolving BLR kinematics in some AGNs over the several-year monitoring baseline. These observations may indicate that the BLR kinematics can change over relatively short timescales. However, it should be noted that there could be systematics when comparing results from different campaigns.  Continuous monitoring plus homogeneous data reduction is best suited for addressing this issue.

Seoul National University AGN Monitoring Project (SAMP) continuously monitored 32 moderate to high luminosity AGNs  from 2016 to 2021.  The long duration and moderate cadence of SAMP light curves allow for constraining the velocity-resolved lags. In addition, the SAMP sample greatly supplements the current velocity-resolved RM sample, particularly at the high-luminosity end. The selection of the SAMP sample and the initial lag measurements of two AGNs are provided in \citet[][Paper \RNum{1}]{Woo19} and \citet[][Paper \RNum{2}]{Rakshit19}, respectively. The final \hbeta\ lags of 32 AGNs based on 6-yr observations are summarized by \citet[][Paper \RNum{3}]{Woo24}, and the \halpha\ lags of six AGNs are reported by \citet[][Paper \RNum{4}]{Cho23}. In the fifth of our paper series, we present the \hbeta\ velocity-resolved lags. 
 
This paper is organized as follows:  \S \ref{sec:observation} briefly summarizes the observation and data reduction. \S \ref{sec:results} presents the \hbeta\ velocity-resolved lags based on multi-year observations, and demonstrates evidence of evolving BLR kinematics by comparing the results in different years.  \S \ref{sec:discussion} combines SAMP velocity-resolved lags with literature measurements  to study connection between BLR kinematics and different AGN properties. \S \ref{sec:conclusion} provides a summary of our findings and conclusions. In this work, we adopt $H_0=72.0$ km s$^{-1}$ Mpc$^{-1}$ and $\Omega_0=0.3$.

\begin{figure*}[htbp]
   \centering
   \includegraphics[width=0.99\textwidth]{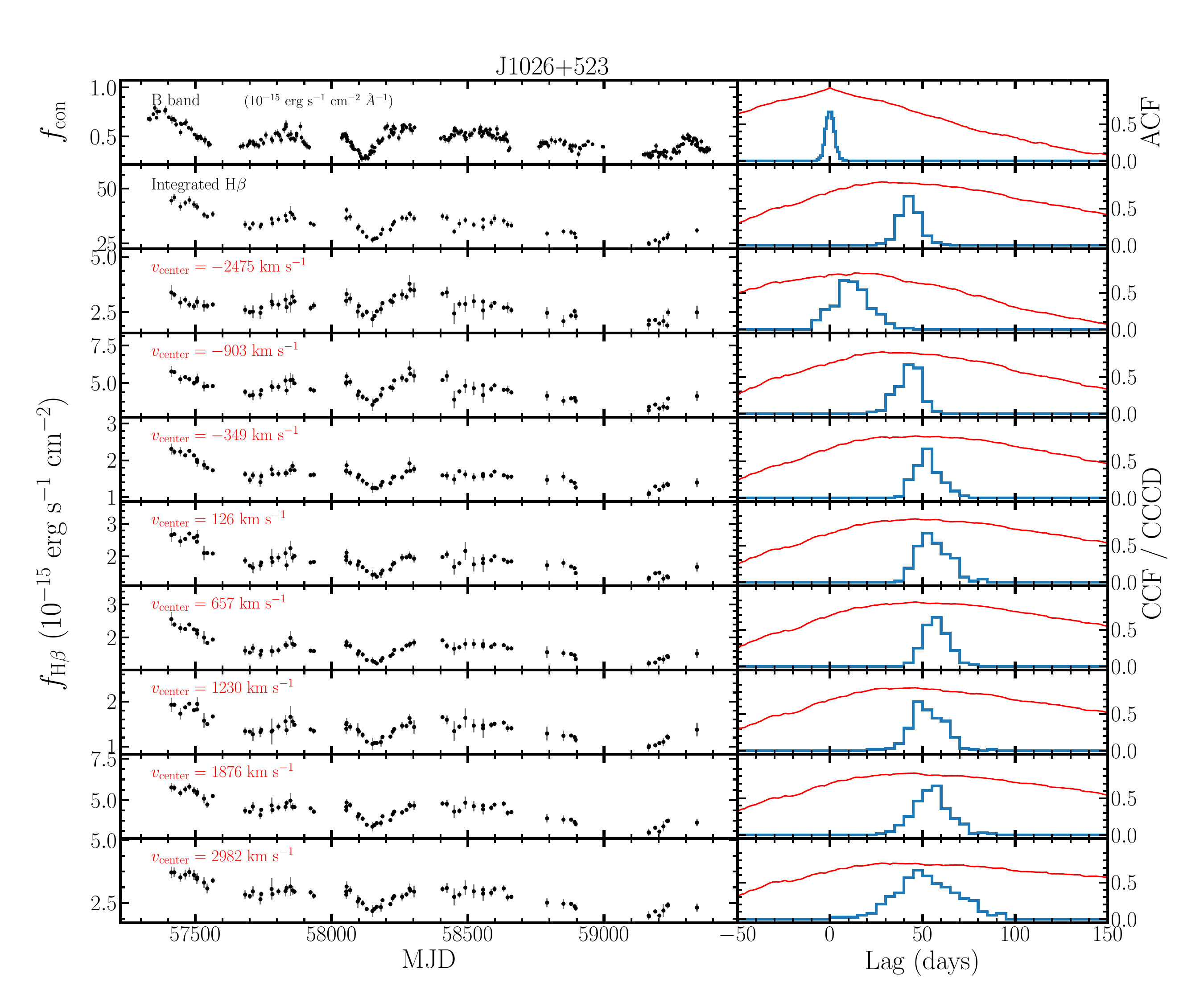} 
   \caption{The velocity-resolved \hbeta\ light curves and the corresponding observed-frame lag measurements for an exemplary object J1026$+$523. On the left, the top panel shows the continuum light curve, while the following panels represent the integrated and velocity-resolved \hbeta\ light curves in different velocity bins, with the center of each bin $v_{\rm center}$ labeled at the upper left corner.  On the right,  each panel displays the CCF (or ACF; red lines)  as well as the CCCD (blue histograms) in the observed frame. }
   \label{fig:LCs} 
\end{figure*}

\section{Observations and Data reduction}  \label{sec:observation}

The initial sample consists of 100 moderate-to-high-luminosity AGNs selected from the MIllIQUAS catalog \citep[Milliquas v4.5 update;][]{Flesch2015} and the Palomar-Green quasar catalog \citep{Boroson92}. Based on a pilot variability test, we select 32 AGNs as our final sample, which spans a redshift range of $z=[0.079, 0.343]$ and a luminosity range of $L_{\rm 5100,AGN}=10^{44.1\sim45.6}$ erg s$^{-1}$. Detailed properties of these targets are summarized in Paper \RNum{3}.  This sample has been continuously monitored both photometrically and spectroscopically from 2015 to 2021.

We carried out photometric observations using several telescopes, including the MDM 1.3 and 2.4 m telescope; the Lemmonsan Optical Astronomy Observatory 1 m telescope; the Lick observatory 1 m telescope; the Las Cumbres Observatory Global Telescope network; and the Deokheung Optical Astronomy Observatory  1 m telescope. The photometry was performed using either $B$/$V$- or $V$/$R$-band filters, depending on the redshift (see Paper \RNum{3}). We followed the standard procedures for data reduction.
Aperture-based photometry was obtained using the {\tt SExtractor} software \citep{BA96} where the initial aperture size was set as three times the seeing size. As described in detail in Paper III, we intercalibrated photometric light curves from different telescopes using the software {\tt PyCALI} \citep{Li14}, which employs a damped random walk model to characterize the variability. The light curves from the MDM 1.3 m telescope were used as the reference because they have the largest number of epochs and are the most evenly distributed over the baseline. The light curves from all other telescopes were aligned to the reference by applying additive and scaling parameters derived from {\tt PyCALI}, with the systematic uncertainty taken into account. For seven objects with relatively lower cadence, we combined our continuum light curves with $g$ band light curves from Zwicky Transient Facility \citep{Bellm19}.

We carried out spectroscopic observations using the Lick Shane 3 m and MDM 2.4 m telescope. As we provided the details of observations and data reduction process in Paper III, here we briefly describe these procedures. We used the Kast double spectrograph for the Lick observations. 
In this work, we utilized only the red side spectrum for studying the \hbeta\ adjacent region. The red side observation employed a 600 lines mm$^{-1}$ grating with a  4$^{\prime\prime}$ slit to minimize the flux loss, providing a spectral resolution R=624. 
We obtained high quality spectra with a typical signal-to-noise ratio (S/N) 15-20.
For the MDM observations we utilized the volume phase holographic blue grism, covering a wavelength range of 3970--6870\AA.
Initially we used a 3$^{\prime\prime}$ slit before 2017 January. Then we used a new customized 4$^{\prime\prime}$ slit in order to secure a consistent setup with Lick observations.

We performed standard spectroscopic data reduction using IRAF and python package {\tt PypeIt}\footnote{\url{https://pypeit.readthedocs.io/en/latest/}} \citep{Prochaska20, Prochaska20b}. The reduced spectra were then aligned based on their \OIII\ profile using the python package {\tt mapspec} \citep{Fausnaugh17}.
The \OIII\ profile was extracted using a fixed window of [4968, 5055] \AA, with the local continuum fitted in the rest-frame windows at [4963, 4968] and [5055, 5060] \AA. The \OIII\ profile was then aligned across different epochs using three parameters: a wavelength shift factor, a flux-scaling factor, and a line-broadening factor. Subsequently, we corrected the small offset between the MDM and Lick light curves, caused by differing site conditions, using closely paired observations. Finally, the absolute flux level was determined by matching the synthetic V band with the photometric V band light curves (R band for two high-z AGNs) using {\tt PyCALI}. 

Based on these calibrated single-epoch spectra, we created the mean and rms spectra which is utilized to generate the velocity bins.  Multi-component fits were performed to decompose the single-epoch spectrum (see Paper \RNum{3}) and the continuum model subtracted emission-line spectrum forms the basis for the following velocity-resolved RM analysis.

In addition, we measured the asymmetry of mean broad \hbeta\ profile using the asymmetry index $A$ \citep{DeRobertis85,Boroson92} defined as:

\begin{equation}
    A = \frac{ \lambda_c(3/4) -  \lambda_c(1/4)}{\Delta \lambda(1/2)}
\end{equation} \label{equ:A}

\noindent where  $\lambda_c(3/4)$ ($\lambda_c(1/4)$) represents the average wavelengths of the two points where the flux reaches 3/4 (1/4) of its peak value, and $\Delta \lambda(1/2)$ is the full-width-at-half-maximum (FWHM) of broad \hbeta. Emission lines with an excess in the red wings exhibit $A<0$, while those with strong blue wings demonstrate $A>0$. The values of $A$ of our sample are provided in Table \ref{tab:BLR-kinematics}. We find that most of our sample show slight red asymmetry, while only three objects exhibit blue asymmetry.

\begin{figure*}[ht]
    \centering
   \includegraphics[width=0.32\textwidth]{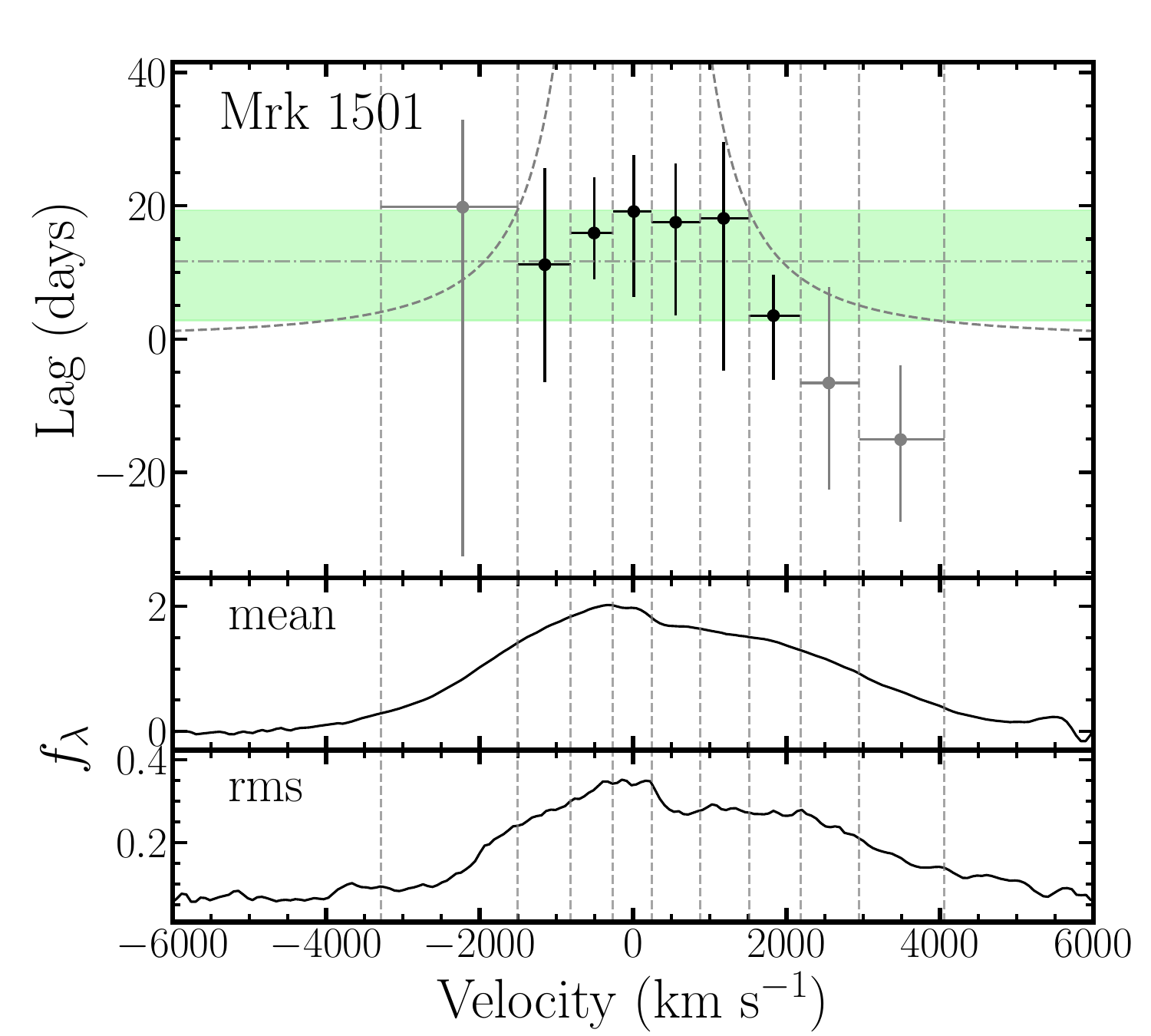}
    \includegraphics[width=0.32\textwidth]{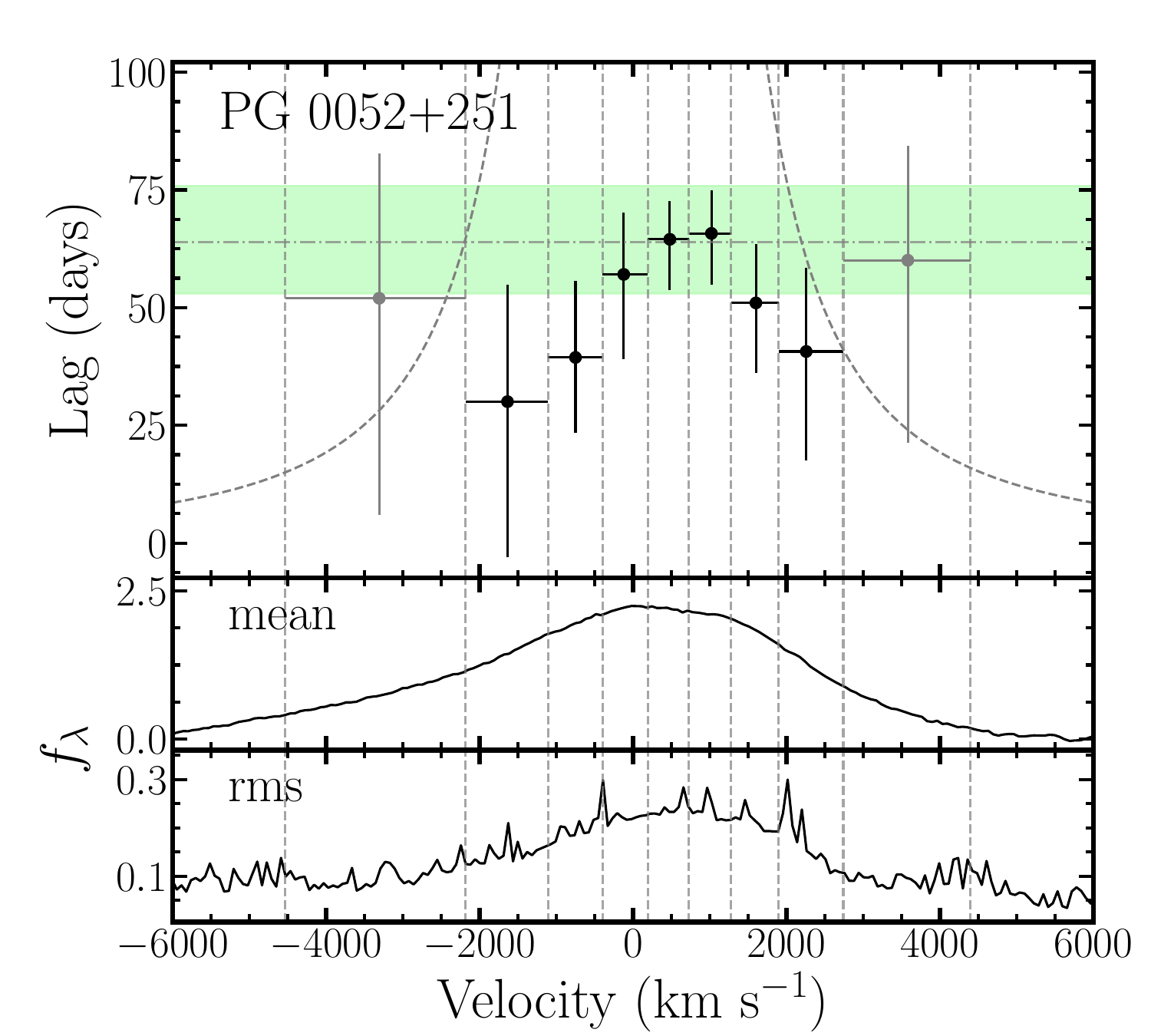}
   \includegraphics[width=0.32\textwidth]{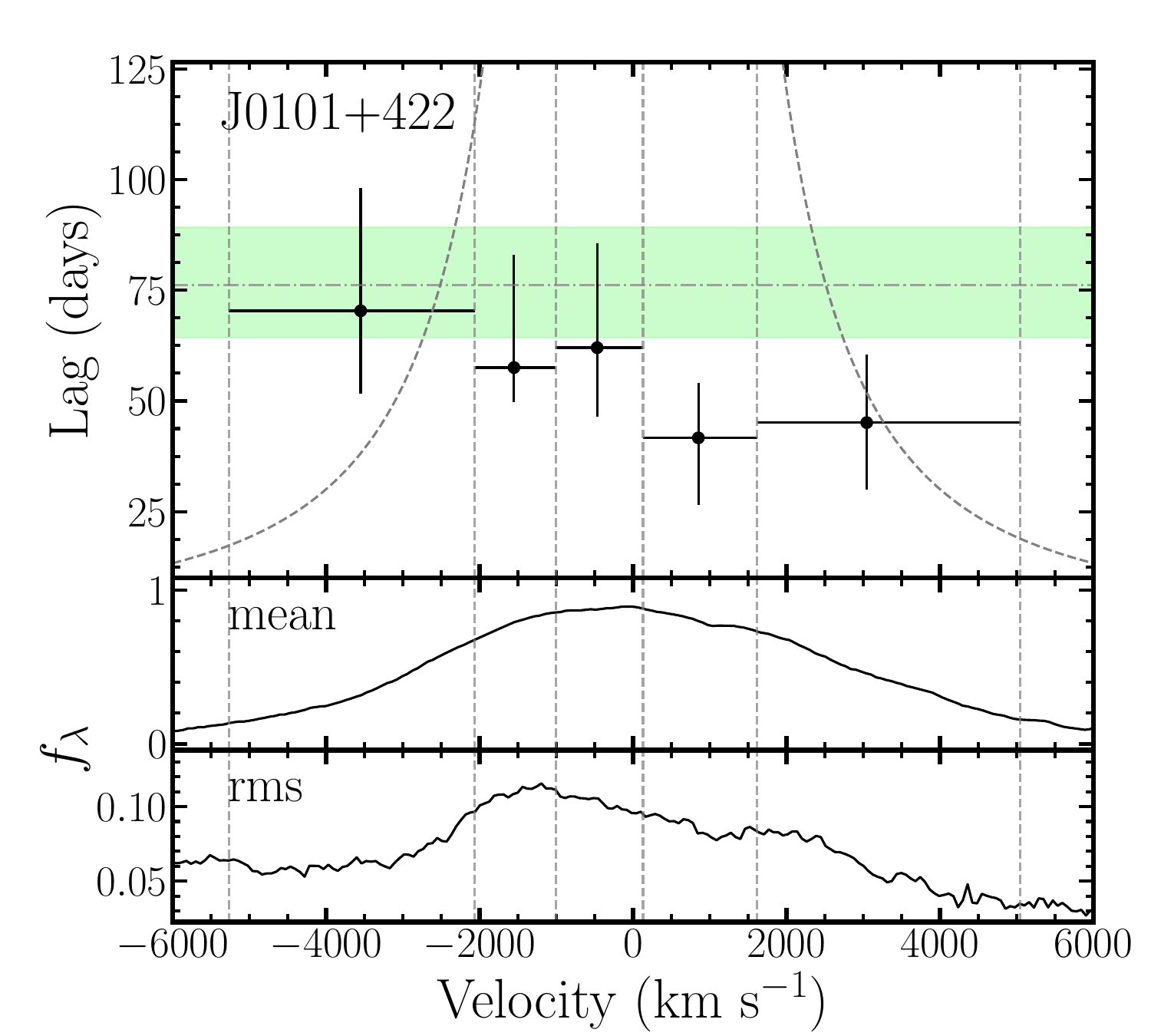}
   \includegraphics[width=0.32\textwidth]{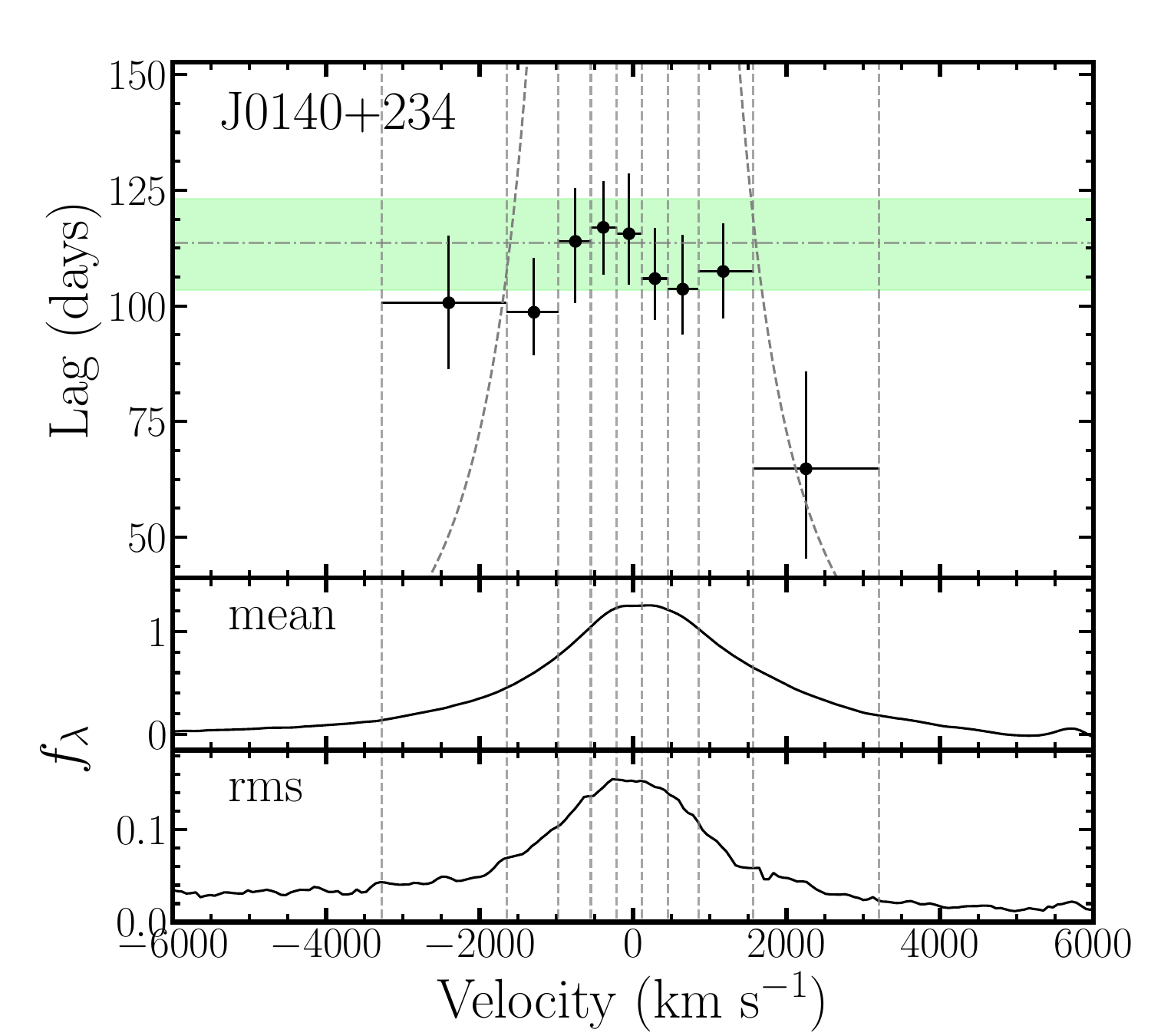}
   \includegraphics[width=0.32\textwidth]{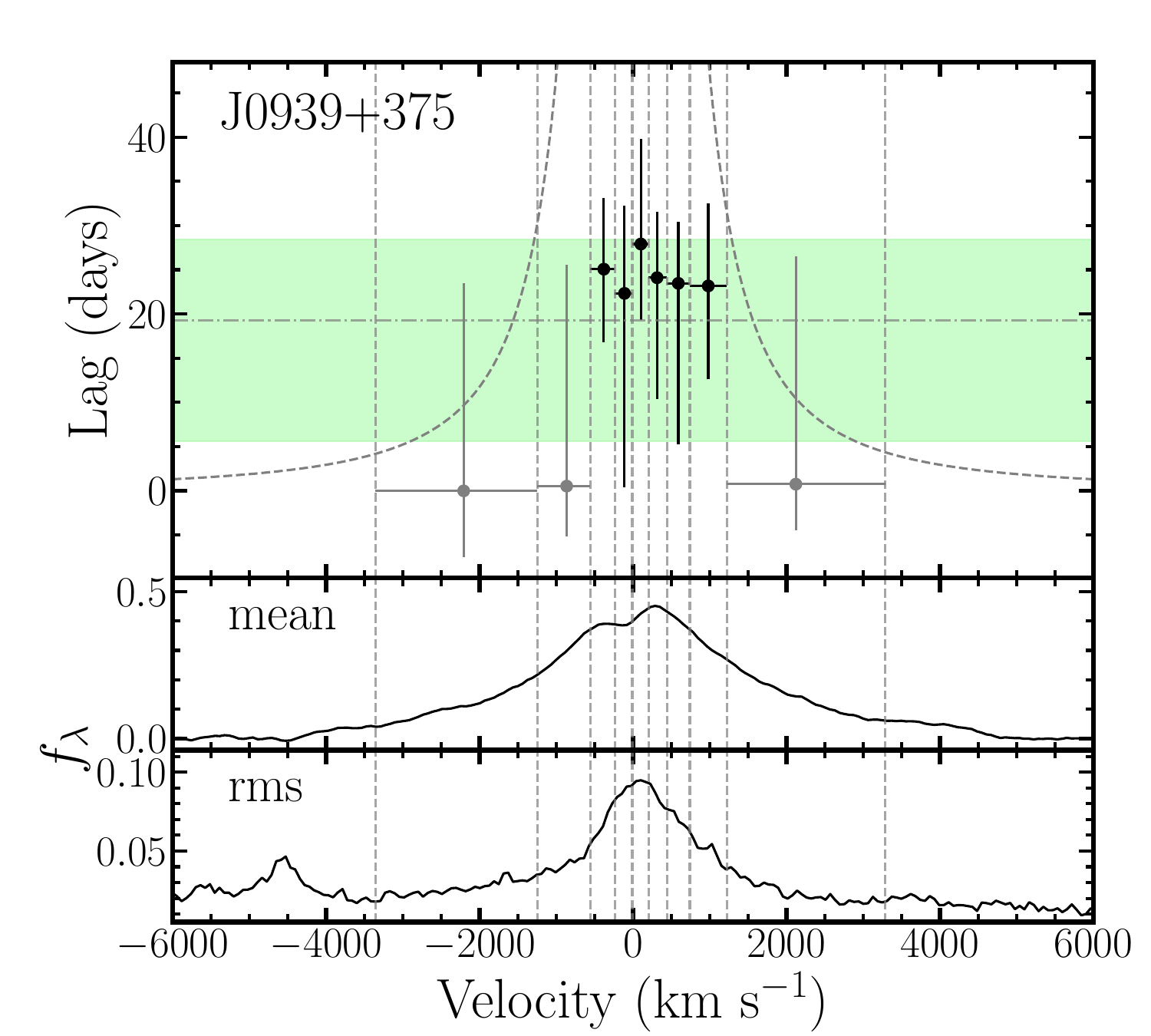}
   \includegraphics[width=0.32\textwidth]{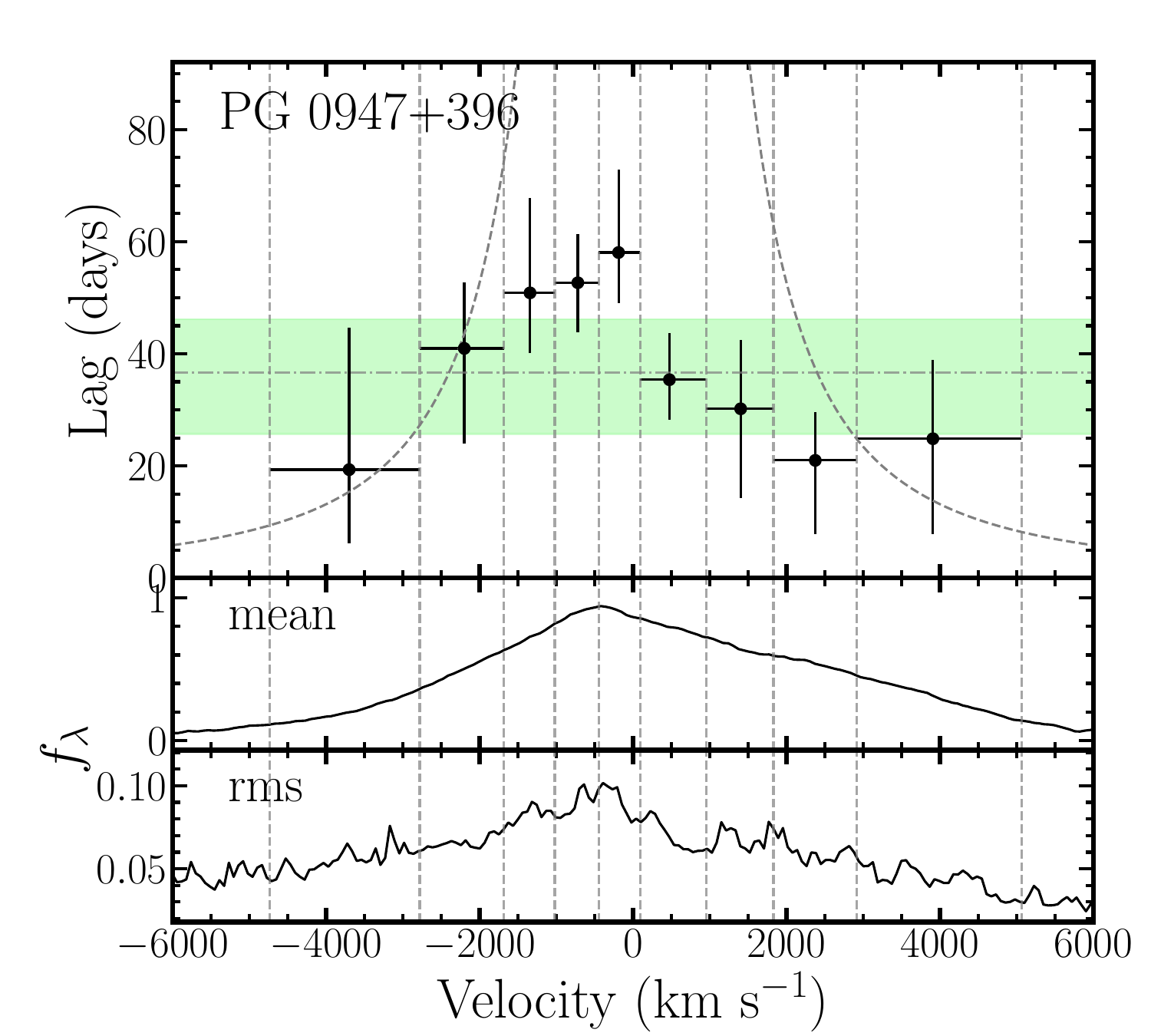}
   \includegraphics[width=0.32\textwidth]{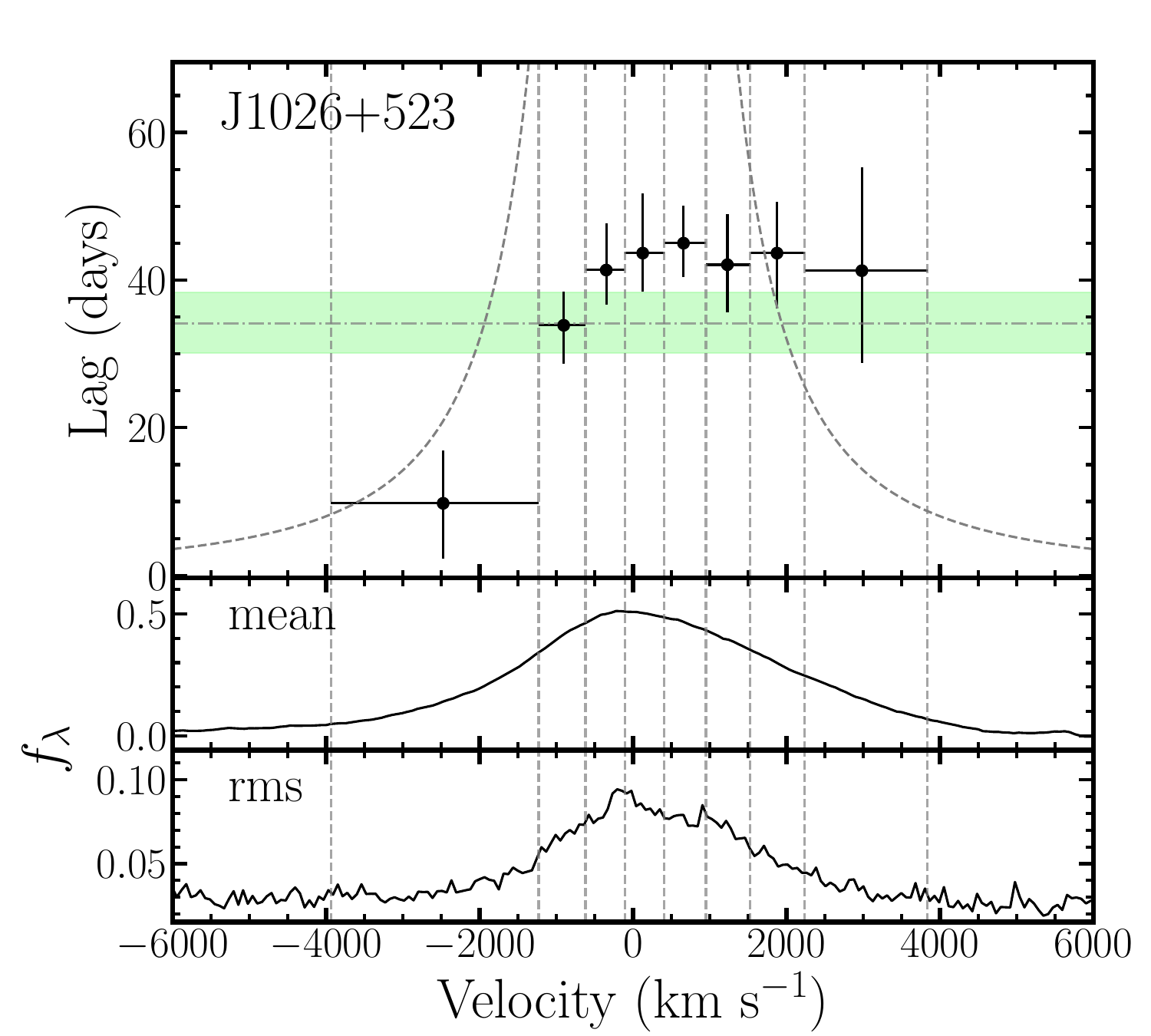}
   \includegraphics[width=0.32\textwidth]{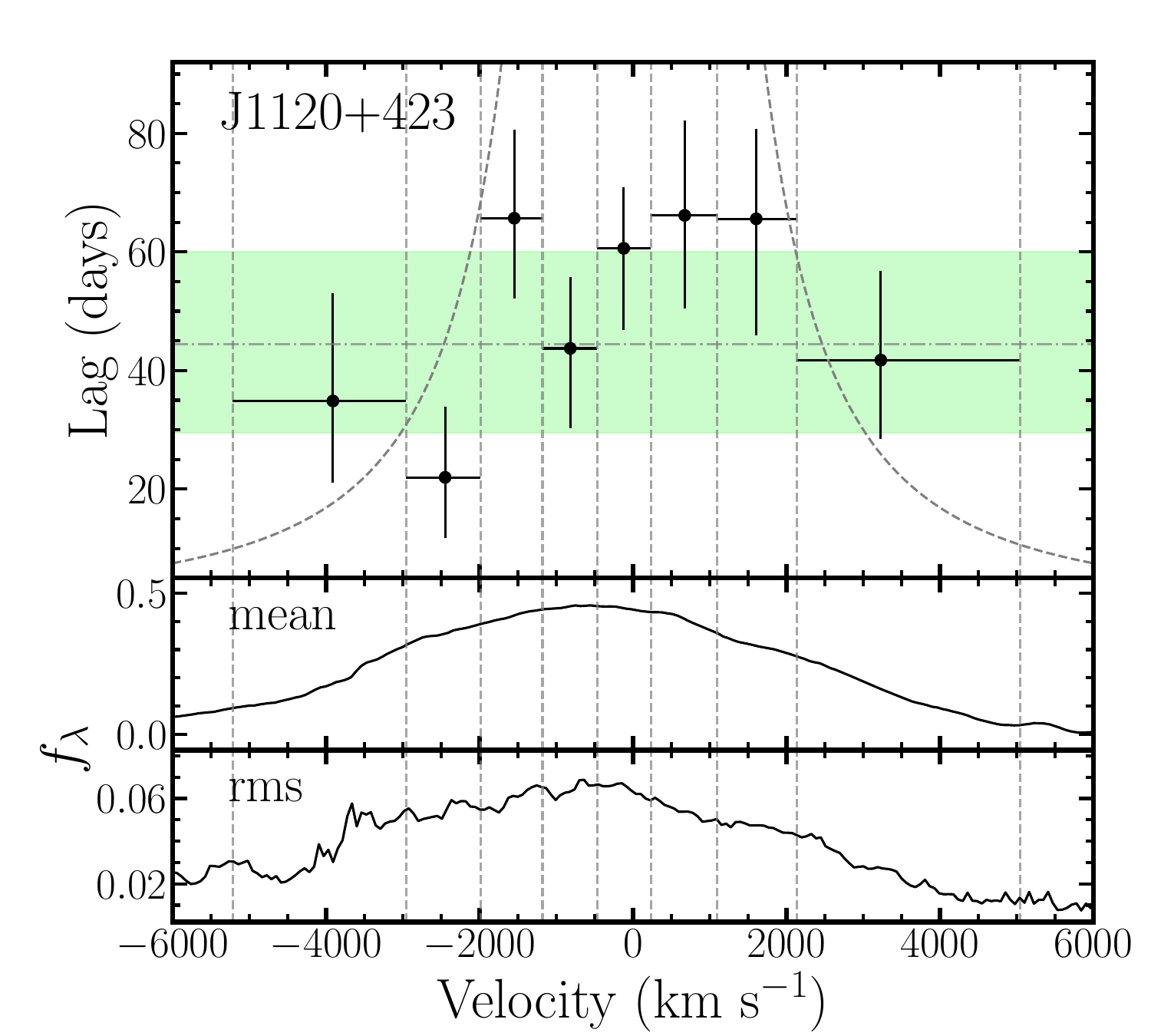}
   \includegraphics[width=0.32\textwidth]{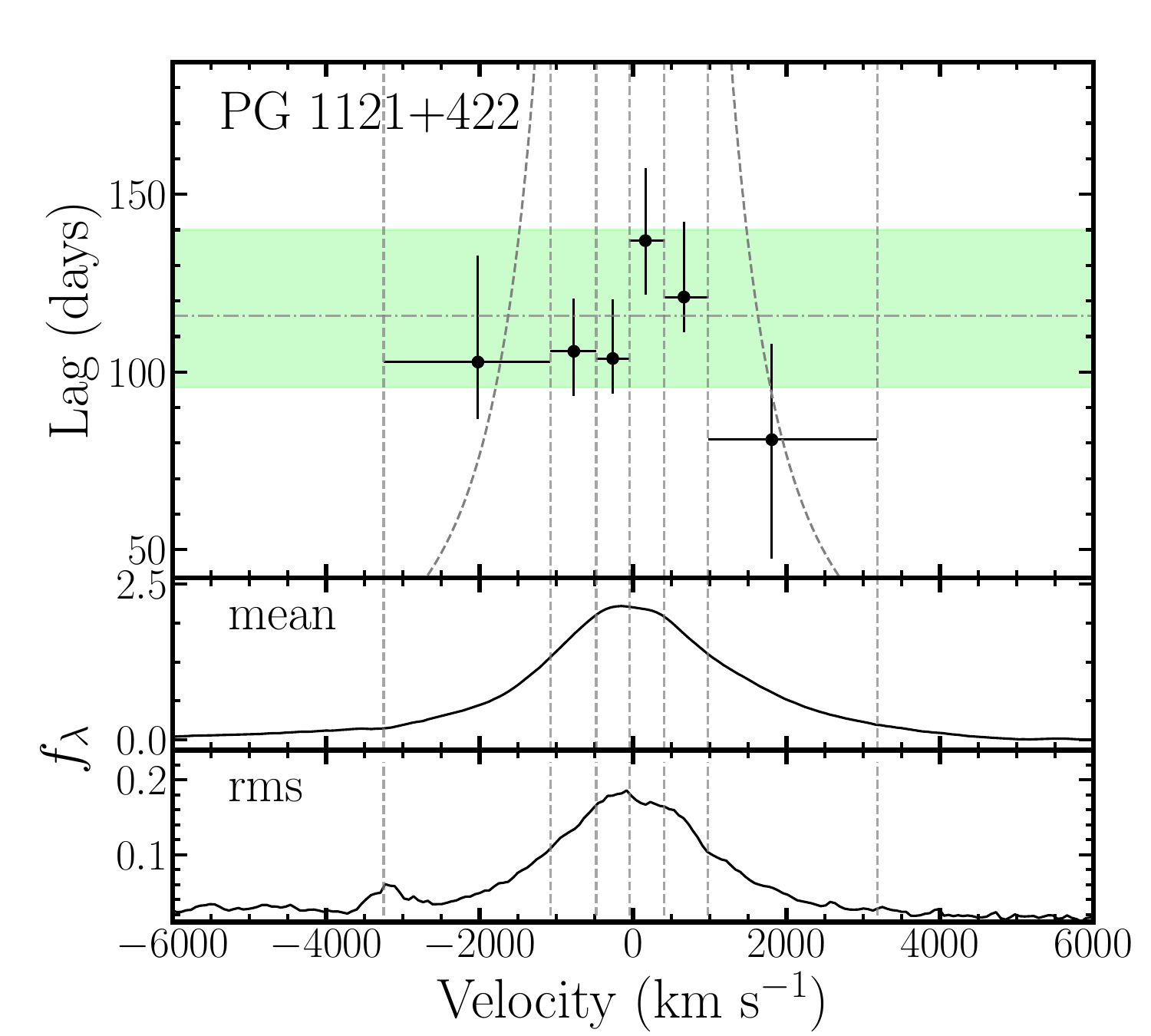}

    \caption{Velocity-resolved lags (top panel) of 20 AGNs from SAMP along with the corresponding mean (middle panel) and rms (bottom panel) spectra in the unit of 10$^{-15}$ erg s$^{-1}$ cm$^{-2}$ ${\rm \AA}^{-1}$. The vertical dashed lines are the edges of velocity bins. The horizontal dotted dashed lines and light green shaded area represent the integrated \hbeta\ lag  and its 1$\sigma$ uncertainty,  respectively (Paper \RNum{3}). Negative lags and lags with small $r_{\rm max}$ are degraded and plotted in grey color. To guide the eyes, we overplot the virial envelope that characterizes a Keplerian, disk-like rotation, with constant BH mass, i.e., $V^2 \times \tau =$ constant.  }
    \label{fig:VRL-1}
\end{figure*}

\addtocounter{figure}{-1}

\begin{figure*}[htp]
    \centering      
      \includegraphics[width=0.32\textwidth]{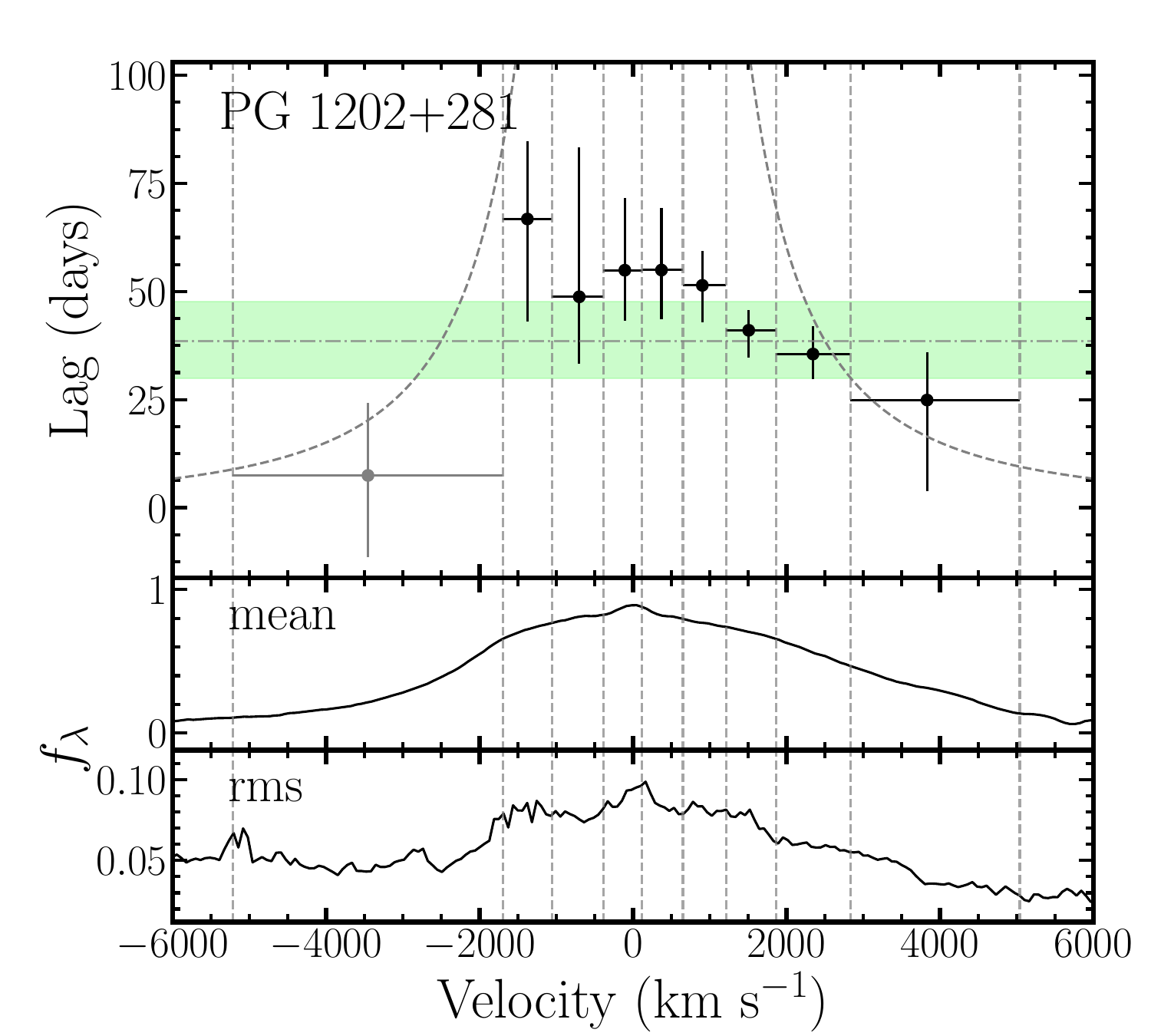}
      \includegraphics[width=0.32\textwidth]{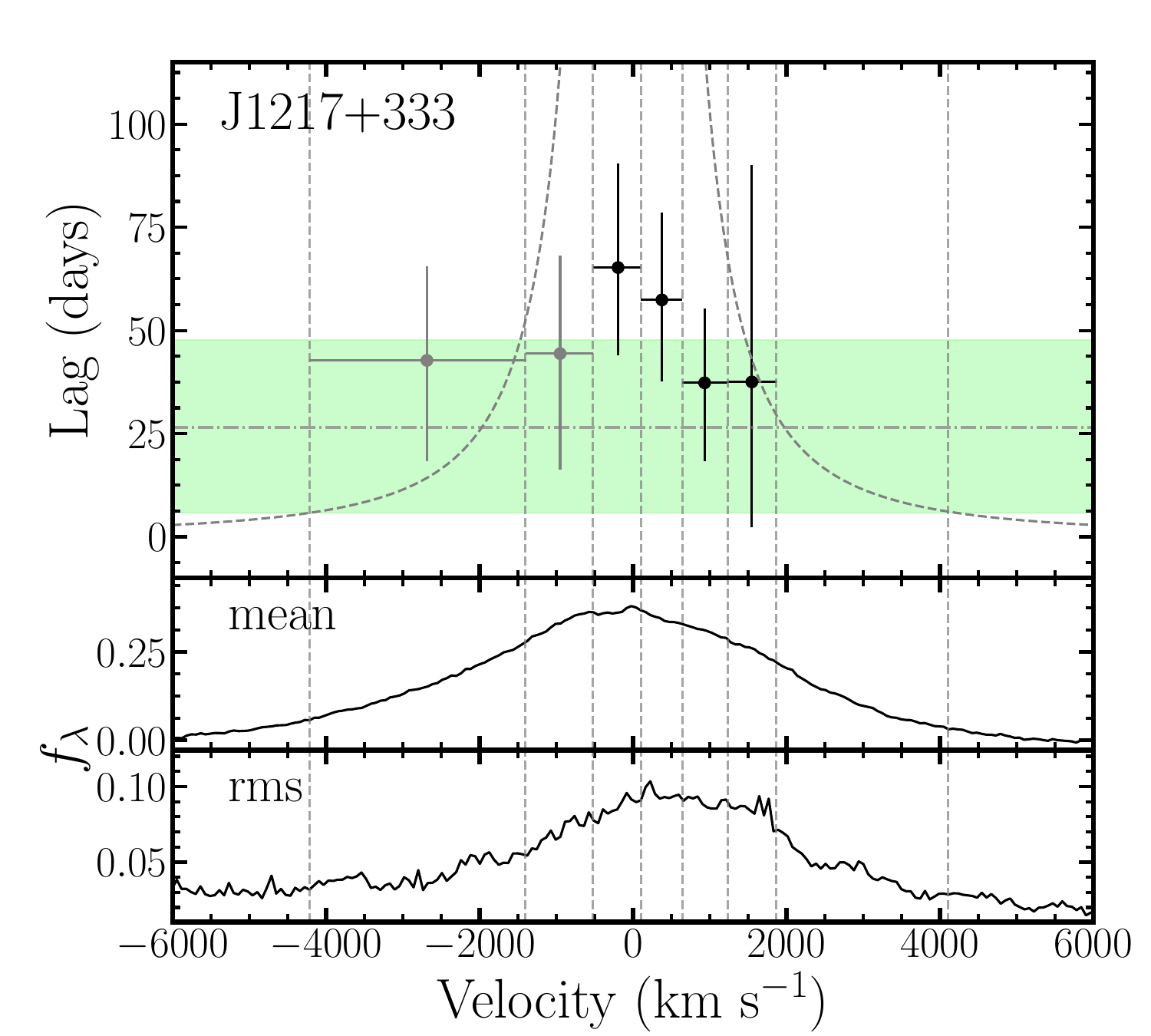}
      \includegraphics[width=0.32\textwidth]{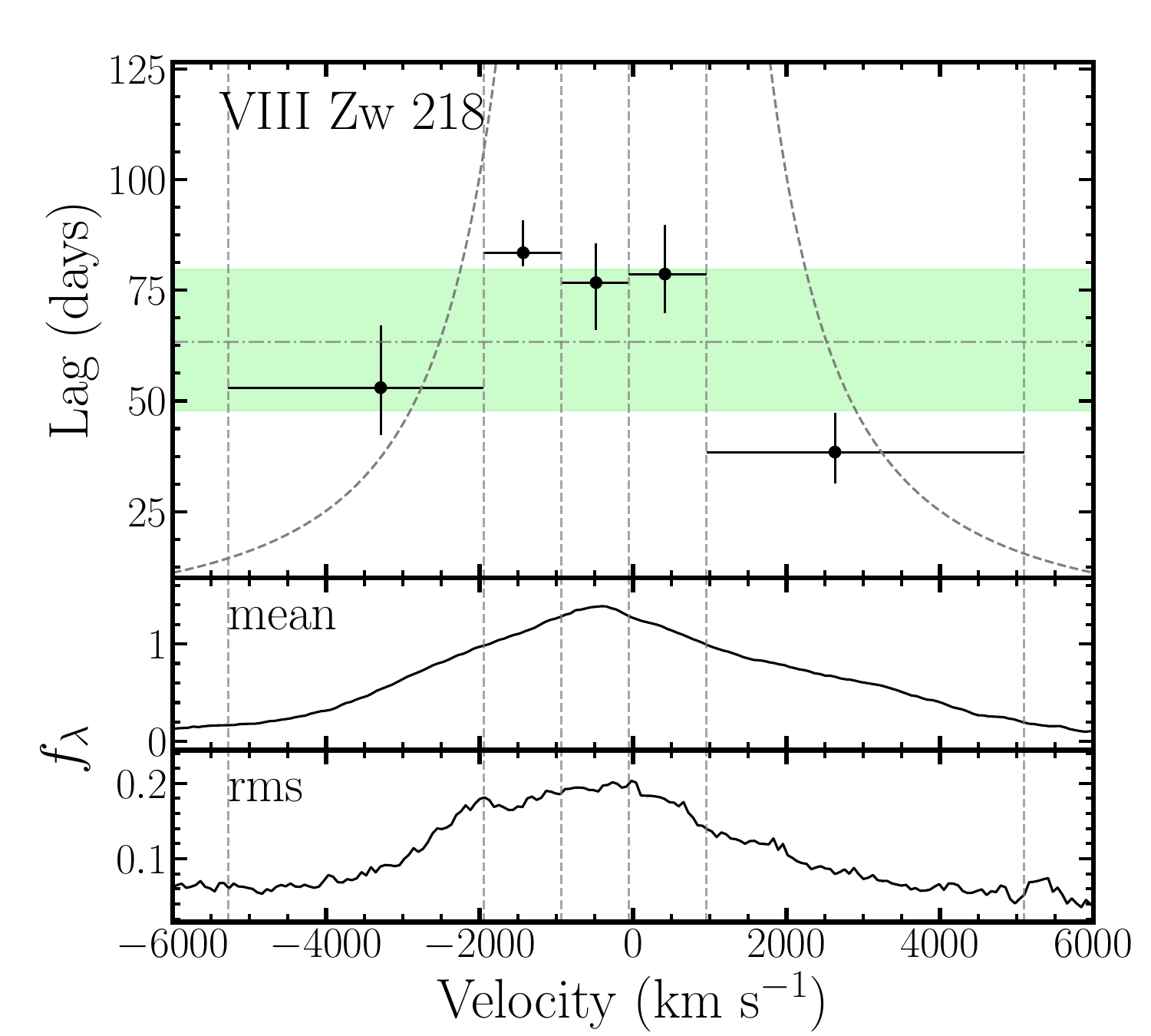}
     \includegraphics[width=0.32\textwidth]{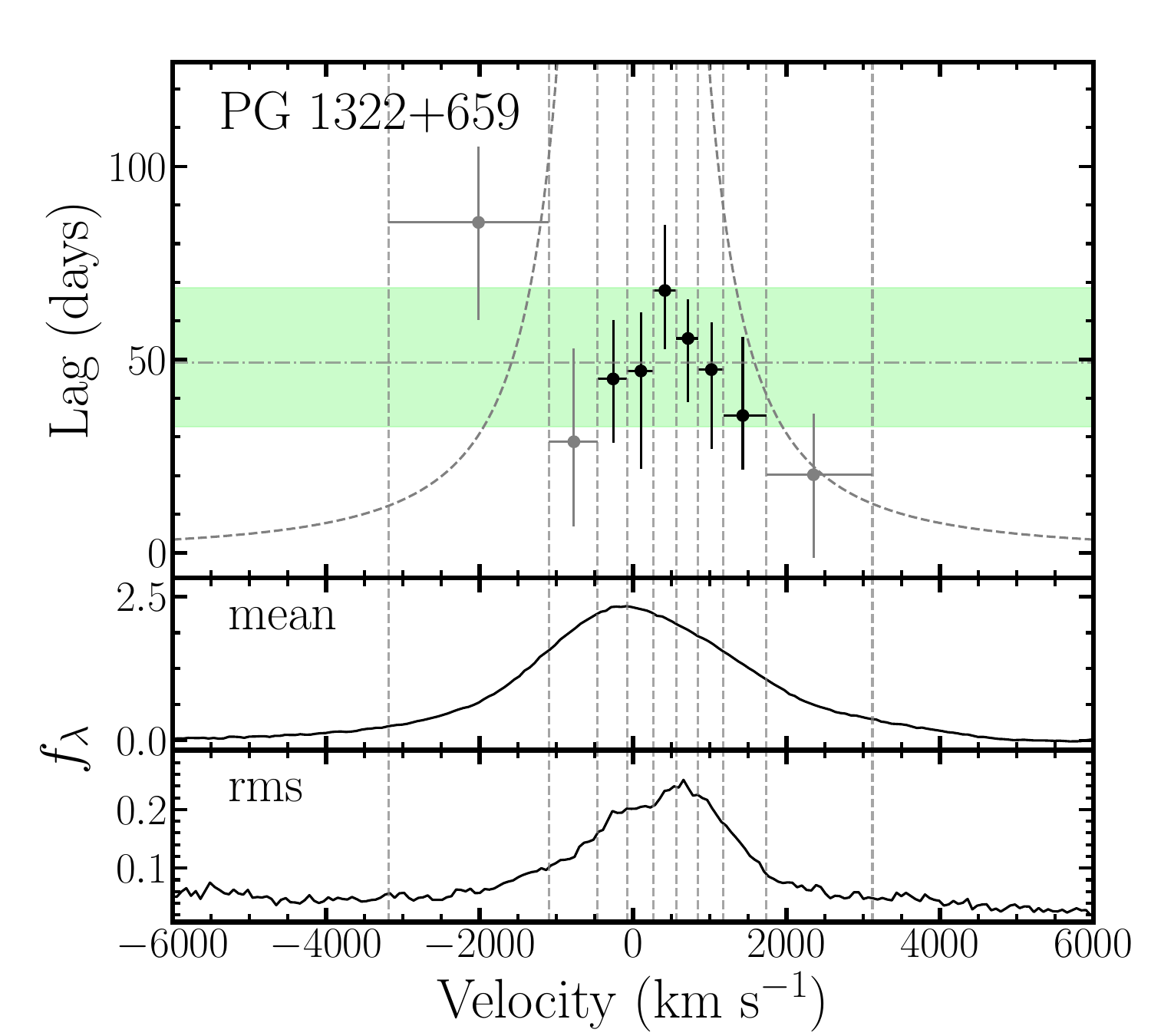}
      \includegraphics[width=0.32\textwidth]{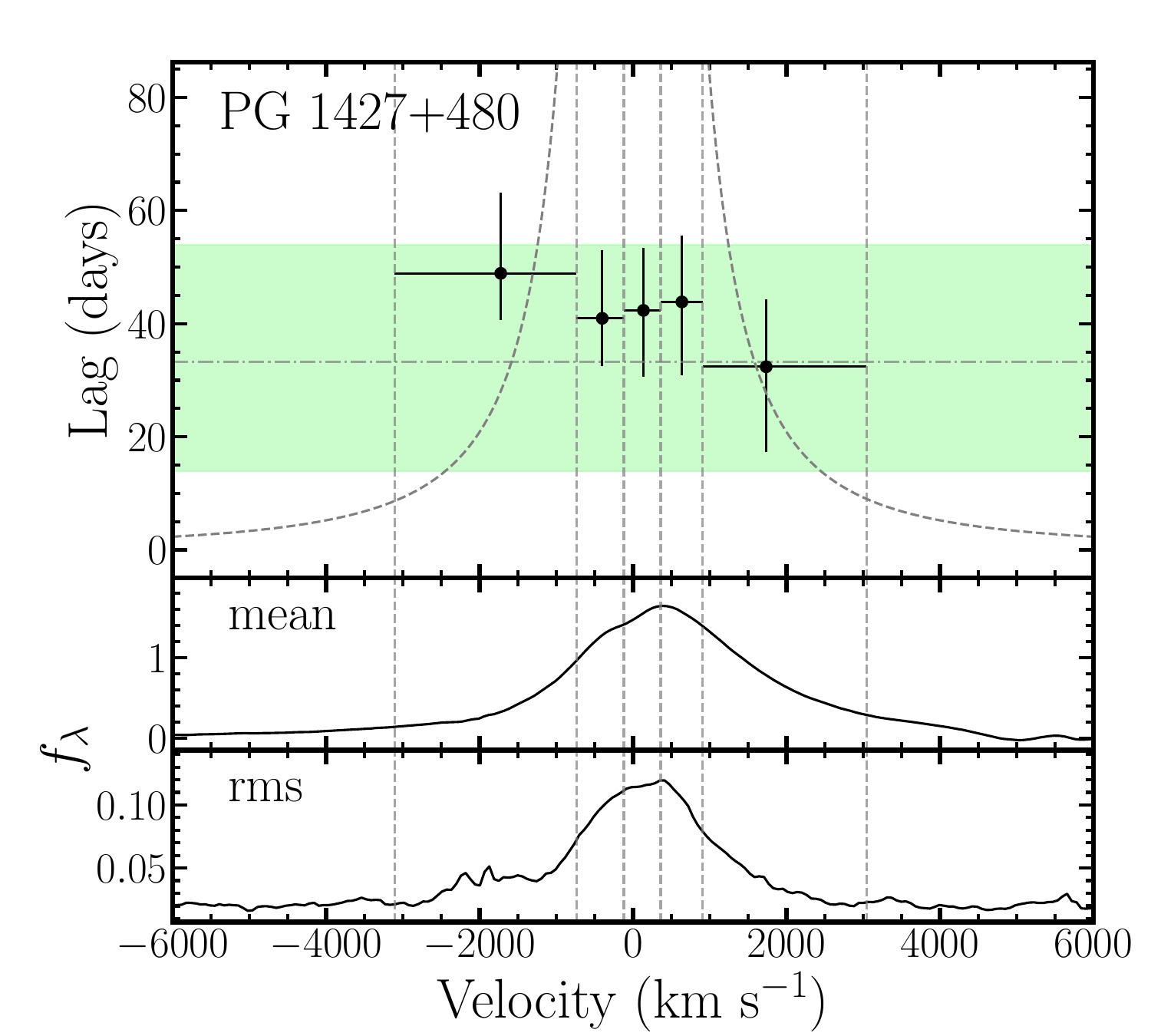}
      \includegraphics[width=0.32\textwidth]{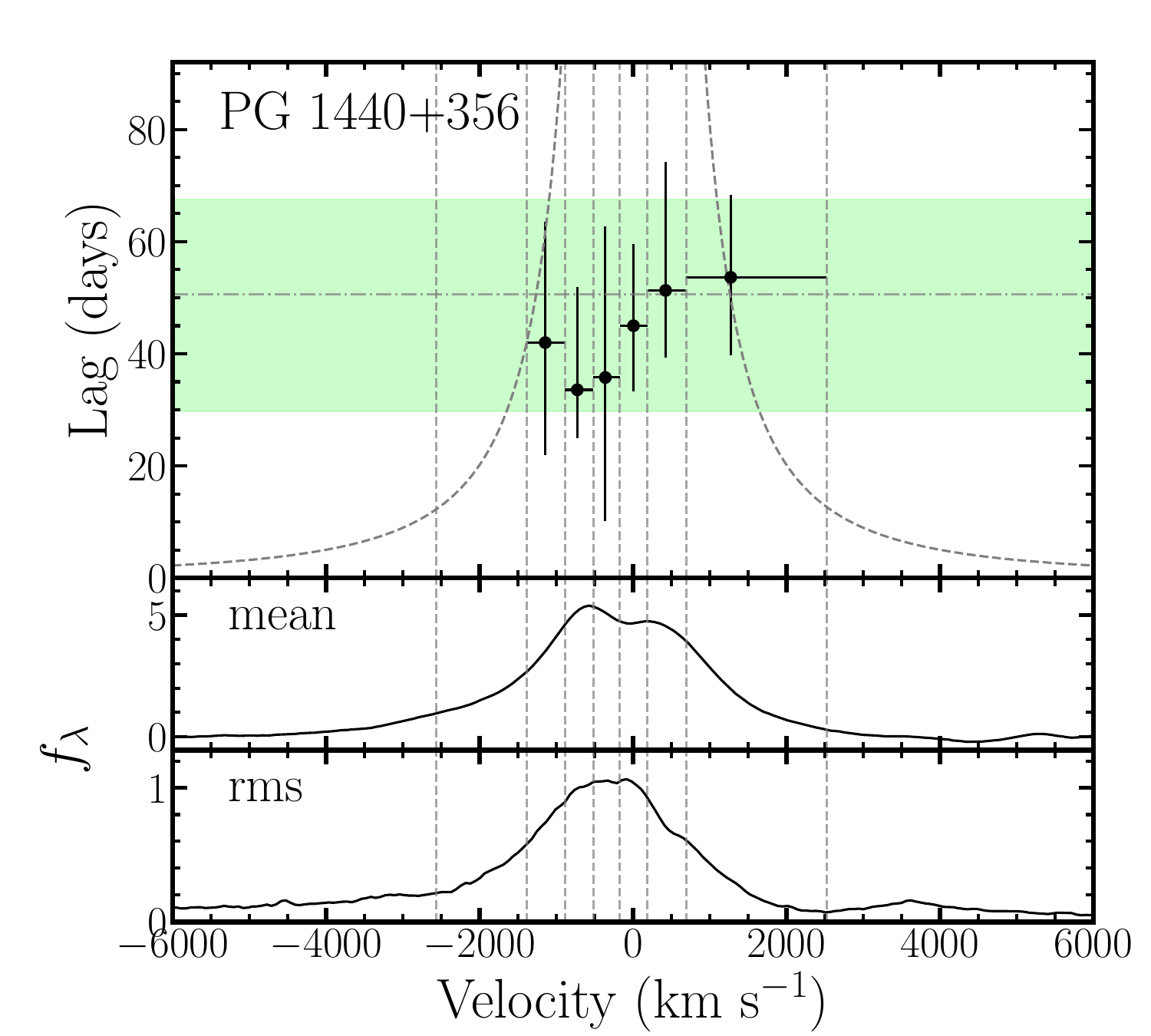}
       \includegraphics[width=0.32\textwidth]{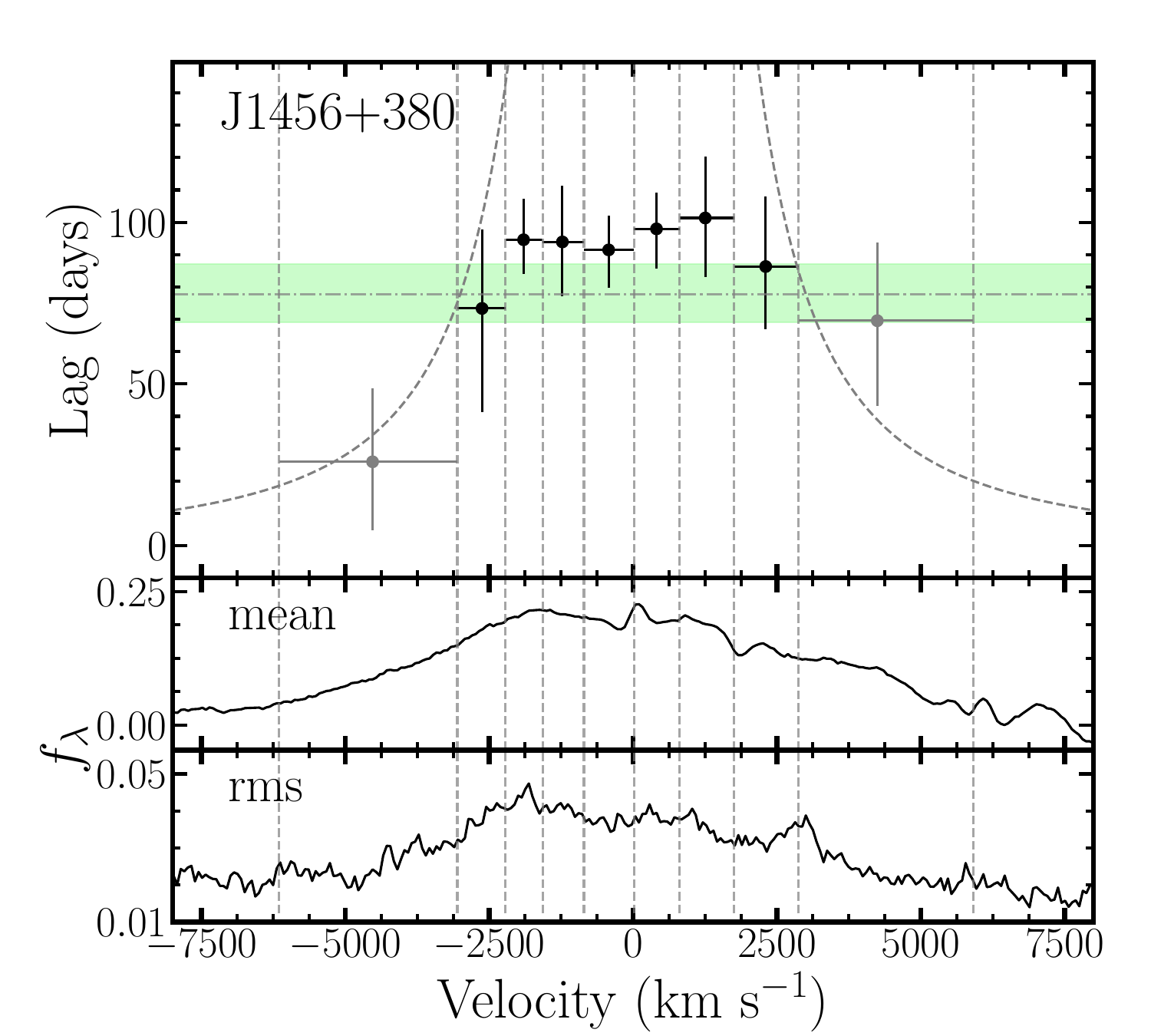}
     \includegraphics[width=0.32\textwidth]{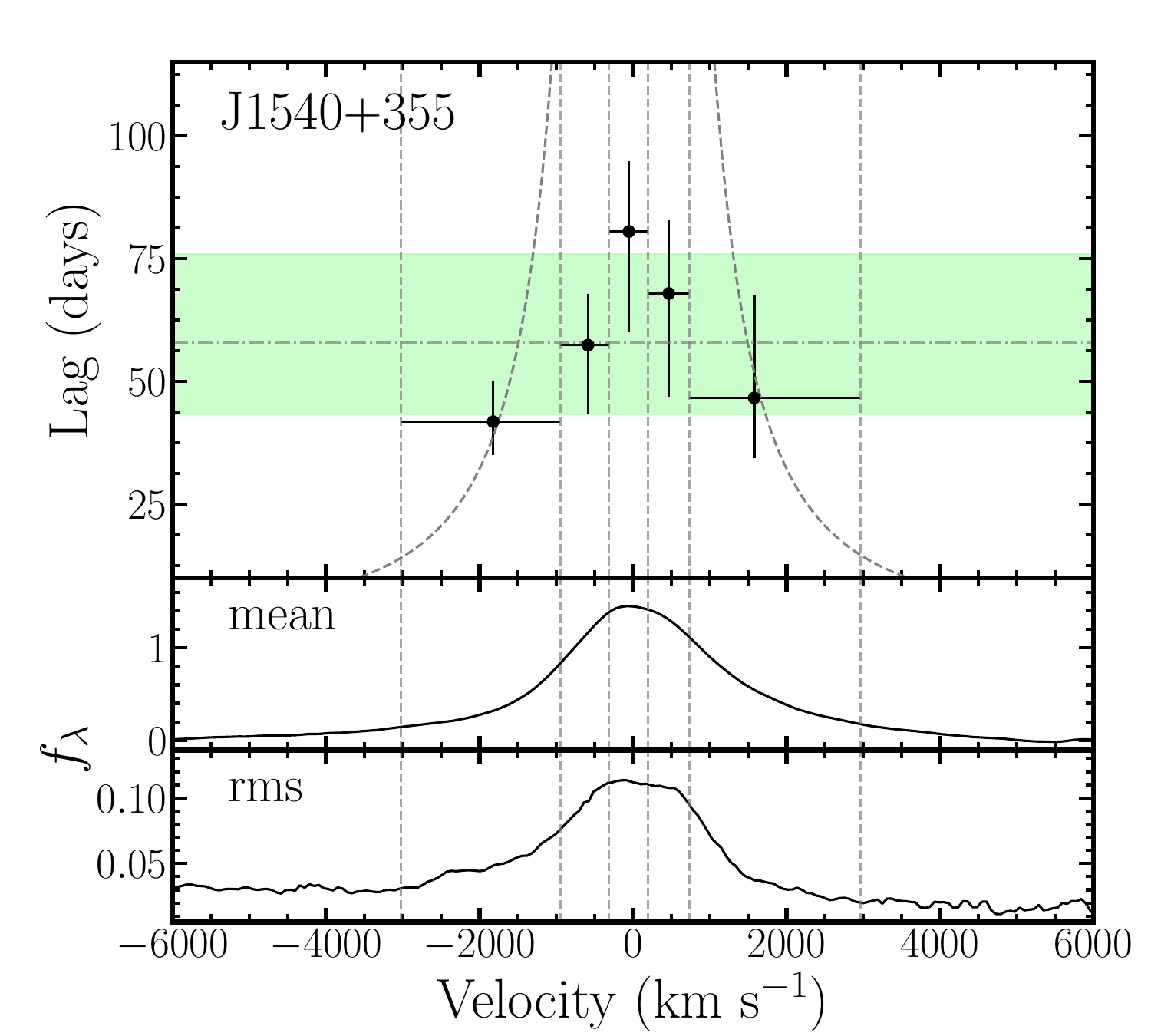}
      \includegraphics[width=0.32\textwidth]{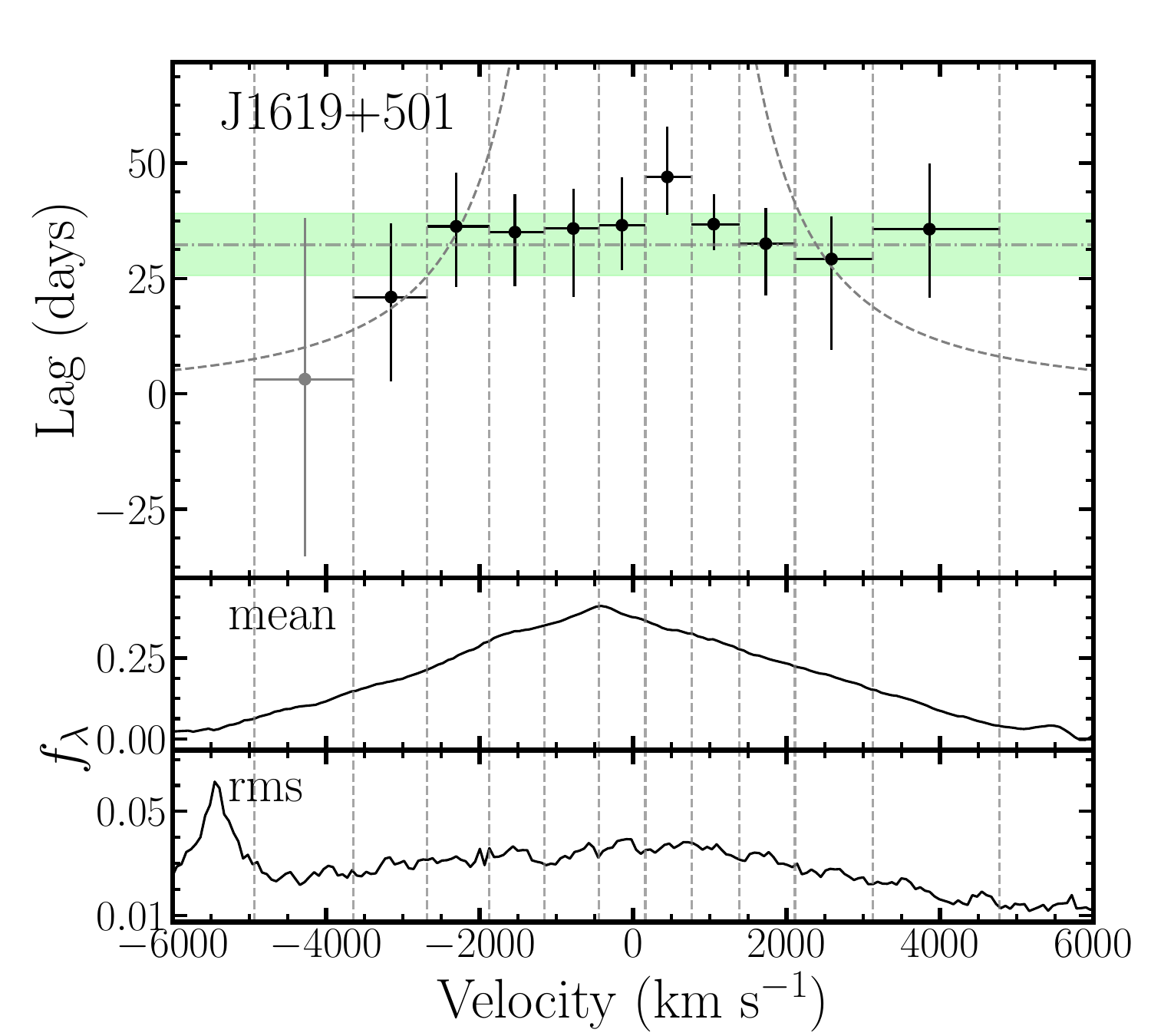}
      \includegraphics[width=0.32\textwidth]{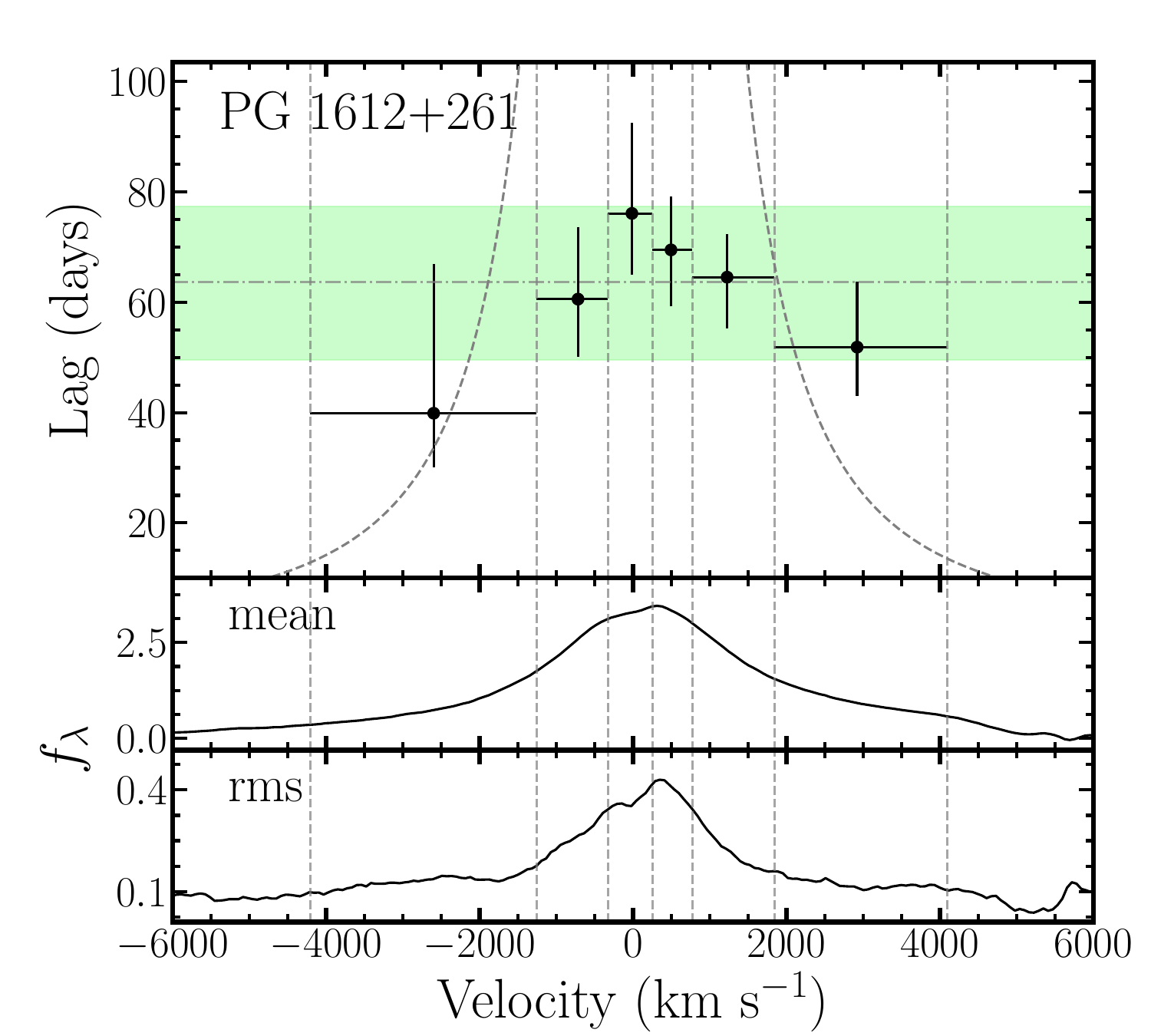}
     \includegraphics[width=0.32\textwidth]{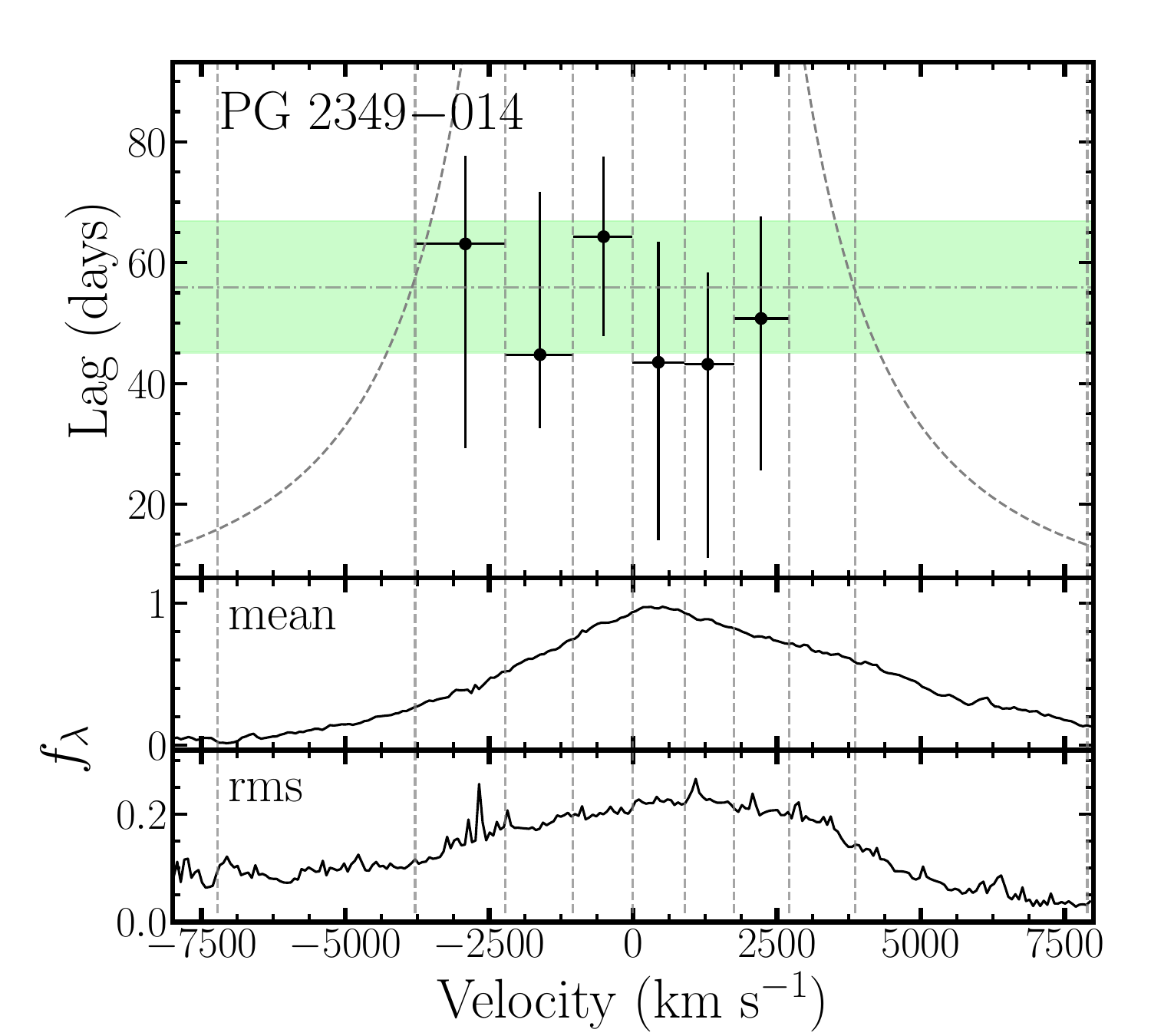}

   \caption{Continued.}
\end{figure*}

\begin{table*}[htbp]
    \centering
    \caption{Velocity-resolved Lag Structures and the Interpretation of BLR Kinematics for 20 SAMP AGNs}
    \label{tab:BLR-kinematics}
    \begin{tabular}{ l c  c c c c }
    \hline\hline
   Object name &  Year &  \multicolumn{3 }{c}{Velocity-resolved lags} & $A$  \\  \cline{3-5}
                       &           &      Visual inspection  &  ${\Delta \tau}_{\rm corr} / \overline{\tau}$    &   $\chi^2_{\rm asy}$  &  \\  
         (1)         &             (2)            &              (3)               &           (4)               &  (5)  & (6) \\      \hline

Mrk 1501            & 2016--2021   &  Ambiguous  & 0.00  & 0.03  & $-0.099$ \\ 
PG 0052$+$251 & 2016--2021   & Center-lagging-wing / red-lagging-blue (outflow-like)   & 0.00   & 1.65   & $0.133$  \\ 
J0101$+$422      & 2016            &  Blue-lagging-red (inflow-like)      & 0.00   & 0.52  & $-0.027$   \\ 
J0140$+$234     & 2016--2021   &  Symmetric   & 0.12   & 1.10  & $-0.011$ \\ 
J0939$+$375     & 2018             &  Ambiguous    & 0.00  & 0.06 & $-0.032$ \\ 
PG 0947$+$396 & 2016--2021   &  Center-lagging-wing / blue-lagging-red (inflow-like)  & 0.11  & 2.02    & $-0.119$\\ 
J1026$+$523     & 2016--2021    & Red-lagging-blue (outflow-like)  & 0.26  & 2.24 & $-0.044$ \\ 
                           & 2016--2017   & Red-lagging-blue (outflow-like)  & 0.00  & 0.73  &  $-0.064$ \\ 
                           & 2018             & Symmetric   & 0.19  & 0.86 & $-0.047$  \\ 
J1120$+$423      & 2016--2021  & Ambiguous  & 0.18  & 0.68  & $-0.048$\\ 
PG 1121$+$422  & 2016--2021  & Center-lagging-wing   & 0.08  & 1.07  & $-0.052$ \\ 
PG 1202$+$281  & 2016--2021   & Blue-lagging-red (inflow-like)  & 0.26  & 2.77  & $-0.083$\\ 
                           &  2016--2017   & Blue-lagging-red (inflow-like) & 0.02   & 1.94  & $-0.095$ \\ 
                            & 2021            & Symmetric   & 0.21   & 1.35  &  $-0.090$ \\ 
J1217$+$333      & 2016--2021    & Ambiguous  & 0.00  & Nan  & $0.041$\\ 
VIII~Zw~218         & 2019--2021  & Symmetric   & 0.26  & 0.56 & $-0.124$ \\ 
PG 1322$+$659  & 2016--2021   & Symmetric (+ redshifted)  & 0.00  & 0.06 & $-0.055$ \\ 
PG 1427$+$480  & 2018--2021   & Ambiguous   & 0.00  & 0.55 & $-0.072$ \\ 
PG 1440$+$356  & 2016--2018   & Red-lagging-blue (outflow-like)   & 0.00  & $0.44$  &  $0.036 $ \\ 
J1456$+$380      & 2016--2021   &  Ambiguous   & 0.00  & $0.14$ & $-0.039$  \\ 
J1540$+$355      & 2016--2021   &  Symmetric   & 0.00  & 0.11  & $-0.057$   \\ 
PG 1612$+$261 &         2021      &  Symmetric & 0.12  & 0.36 & $-0.074$ \\ 
J1619$+$501      & 2016--2021   & Ambiguous    & 0.00  & 0.29 &  $-0.049$ \\
PG 2349$-$014  & 2018--2021   &  Ambiguous & 0.00   & 0.44  & $-0.115$ \\ \hline
\multicolumn{6}{p{16cm}}{{\bf Notes.} Column (1): Object names in the order of increasing R.A. Column (2):  the year(s) of observation on which our velocity-resolved lags are based.  Column (3): visual inspection results of the velocity-resolved structure.  Column (4): ${\Delta \tau}_{\rm corr} / \overline{\tau}$ represent the excess variance of lags across velocity bins (see \S3.1). Objects with significant resolved lags have ${\Delta \tau}_{\rm corr} / \overline{\tau}>0$.  Column (5): the statistic used to quantify the asymmetry of the velocity-resolved lags (see \S3.1). 
Larger $\chi^2_{\rm asy}$ values indicate stronger difference between lags in blue and red velocity bins. Column (6): the asymmetry index $A$ (see Equation \ref{equ:A}) of the broad \hbeta\ measured from the mean spectra. $A<0$ indicates flux excess in red wings while $A>0$ indicates flux excess in blue wings. }  
    \end{tabular}

\end{table*}

\section{Results} \label{sec:results}
\subsection{Velocity-resolved Lags from Multi-year Observations} \label{sec:vrl-6yr}

In this section, we present the \hbeta\ velocity-resolved lags for 20 AGNs from SAMP sample. These objects were selected from the 25 AGNs with good-quality \hbeta\ lags measured by Paper \RNum{3}.  We imposed additional selection criteria, requiring that the median cadence of the \hbeta\ light curves be smaller than 20 days and the integrated \hbeta\ lag 1$\sigma$ exceeds  zero.  

Figure \ref{fig:LCs} displays an example of \hbeta\  velocity-resolved light curves, along with the continuum and integrated \hbeta\ light curves.  The velocity bins are defined to contain equal flux in the rms spectra,  and the leftmost and rightmost boundaries are set as the flux extraction windows of integrated \hbeta\ (see Table 3 in Paper \RNum{3}).  We also tested defining the bins using mean spectra,  which yielded consistent  results. In some cases, we found a small offset between Lick and MDM light curves. To remove this offset, we performed additional calibration between the MDM and Lick light curves for individual velocity bins using {\tt PyCALI}. 

The lags are measured based on the Interpolated Cross Correlation Function \citep[ICCF;][]{Peterson98, Sun18}, with the ICCF centroids adopted as the lag measurements.  A lag searching window of [$-360$, $360$] days is utilized, which is sufficient based on the integrated lags reported in Paper \RNum{3}. We adopt the same alias removal procedure following the method outlined in Paper \RNum{3}  to clean the cross-correlation centroid distribution (CCCD). The linear detrending procedure \citep[e.g.,][]{Welsh99} is applied for one AGN, PG 1202+281 (see Paper \RNum{3} for details).  Negative lags,  as well as lags of low correlation strength (e.g., the maximum correlation coefficient $r_{\rm max}\leq0.5$), are excluded from subsequent analysis. One additional object (J1540+355) was removed because the ICCF centroid produces a poorer match between the shifted continuum and \hbeta\ light curves for this object compared to alternative methods (e.g., JAVELIN). Most of these velocity-resolved lags were derived using the full six years of data. For a few sources, we discarded parts of the light curves with lower cadence and derived the lags using remaining years.

Figure \ref{fig:VRL-1}  exhibits velocity-resolved lags for  20 AGNs alongside their mean and rms spectra. As can be seen from the figure, some AGNs exhibit clear differences in the lags across different velocity bins, while other objects show no clear variation or trends given the lag uncertainty. We perform visual inspection to examine whether there are clear trends, and further classify the observed trends into symmetric/center-lagging-wing, red-lagging-blue, and blue-lagging-red patterns. To guide our eyes, we overplot the virial envelope in Figure \ref{fig:VRL-1}, which illustrates the case of Keplerian, disk-like rotation, with the BH mass being a constant, i.e., $V^2 \times \tau = $ constant. We adopt the integrated \hbeta\ lag and the velocity dispersion from mean spectra as the normalization of the virial envelope. Using velocity dispersion of rms spectra provides similar results.   The results of our visual inspection are summarized in Table \ref{tab:BLR-kinematics}, and the detailed discussion for each AGN is provided in the Appendix \ref{sec:comments}.

Our visual inspection suggests clear structure for 12 AGNs, among which 8 objects show symmetric/center-lagging-wing structures, 2 objects exhibit blue-lagging-red (inflow-like) structures, and 2 objects display red-lagging-blue  (outflow-like) structures.  
For three AGNs (PG 0947+396, PG 0052+251, and PG 1121+422), though some asymmetry is observed, the main characteristic is a center-lagging-wing structure.
Another object (PG 1322+659) exhibits a shifted peak of the velocity-resolved lags, but the structure is symmetric,  which may indicate a net movement of the BLR or the presence of anisotropic emission in the BLR.

We  investigate the use of quantitative criteria to verify our visual inspection.  
We calculate the  excess variation of lags across the velocity bins, i.e., ${\Delta \tau}_{\rm corr} / \overline{\tau}$, where ${\Delta \tau}_{\rm corr}$ is defined as $\sqrt{\sum_{i}{(\tau_i-\overline{\tau})^2 - \tilde{\sigma}_{\rm err}^2}}$, and $\tau_i$, $\overline{\tau}$, and $\tilde{\sigma}_{\rm err}$ represent the lag in the $i$th bin, the average lag weighted by each bin’s lag uncertainty, and the median lag uncertainty, respectively. If the variance term is smaller than the uncertainty term, we set ${\Delta \tau}_{\rm corr} / \overline{\tau}=0$.  Our analysis reveals that 7 out of the 12 AGNs show ${\Delta \tau}_{\rm corr} / \overline{\tau}>0$, confirming their significant velocity-resolved lags. It indicates that our visual inspection is generally robust. For the remaining 5 AGNs, although their lag uncertainties are relatively large so that their ${\Delta \tau}_{\rm corr} / \overline{\tau}=0$, we do see clear and consistent trends across the velocity bins for two objects (PG 0052+126 and J1540+355). On the other hand, for J1120+423, while it shows ${\Delta \tau}_{\rm corr} / \overline{\tau}>0$, the structure of velocity-resolved lags are ambiguous and the kinematics are indeterminate.   

We also calculate the reduced $\chi^2$ to test for  asymmetry, denoted as $\chi^2_{\rm asy}$. The $\chi^2$ is calculated as $\chi^2 = \sum_{i}{\frac{(\tau_{\rm i,blue} - \tau_{\rm i,red})^2}{\sigma_{\rm i,blue}^2 + \sigma_{\rm i,red}^2}}$, where $\tau_{\rm i,blue}$ ($\sigma_{\rm i,blue}$) and $\tau_{\rm i,red}$ ($\sigma_{\rm i,red}$)  represent the lag (lag uncertainty) of the $i$th pair of the approximately matched blue and red velocity bins, respectively. The $\chi^2_{\rm asy}$ is then calculated as $\chi^2$ divided by $N-1$, where $N$ is the number of bin pairs. 
Larger $\chi^2_{\rm asy}$, e.g., $\chi^2_{\rm asy} \gtrapprox 2$,  indicate more significant asymmetry. We find that three objects  have $\chi^2_{\rm asy} \gtrapprox 2$, which are also identified as asymmetric by our visual inspection (including PG0947+396). For J0101+422 and PG 1440+356, although they do not meet the quantitative criteria due to large lag uncertainties, consistent trends are observed. Consequently, we classify J0101+422 and PG 1440+356 in the inflow and outflow category, respectively.

In summary, we find that 8 out of 20 AGNs exhibit symmetric/center-lagging-wing velocity-resolved lags, consistent with disk-like rotation, 2 AGNs show inflow-like characteristics, and 2 AGNs  display outflow-like signatures. For the remaining 8 AGNs with ambiguous trends, we leave the BLR kinematics interpretation as indeterminate.

We investigated the connection between the structure of velocity-resolved lags and and line profile asymmetry. No clear connection between asymmetry and kinematics is found. For the object with the strongest blue asymmetry in our sample (i.e., PG 0052+251; $A > 0.1$; Table \ref{tab:BLR-kinematics}), the inferred kinematics are disk-like or outflow-like. The other two objects with $A > 0$ exhibit outflow-like (PG 1440+356) and ambiguous (J1217+333) kinematics. For the three objects with the strongest red asymmetry ($A < -0.1$), the inferred kinematics incldue disk-like/inflow-like (PG 0947+396), disk-like (VIII Zw 218), and ambiguous (PG 2349$-$014).
On the other hand, the two outflow-like objects display asymmetries in opposite directions: J1026+523 has a slight red asymmetry, while PG 1440+356 shows a slight blue asymmetry. The two inflow-like objects are either nearly symmetric (J0101+422) or slightly red asymmetric (PG 1202+281).

\subsection{Evidence of BLR Kinematics Evolution} \label{sec:changes_kinematics}

\begin{figure*}[htbp]
    \centering
               \includegraphics[width=0.45\textwidth]{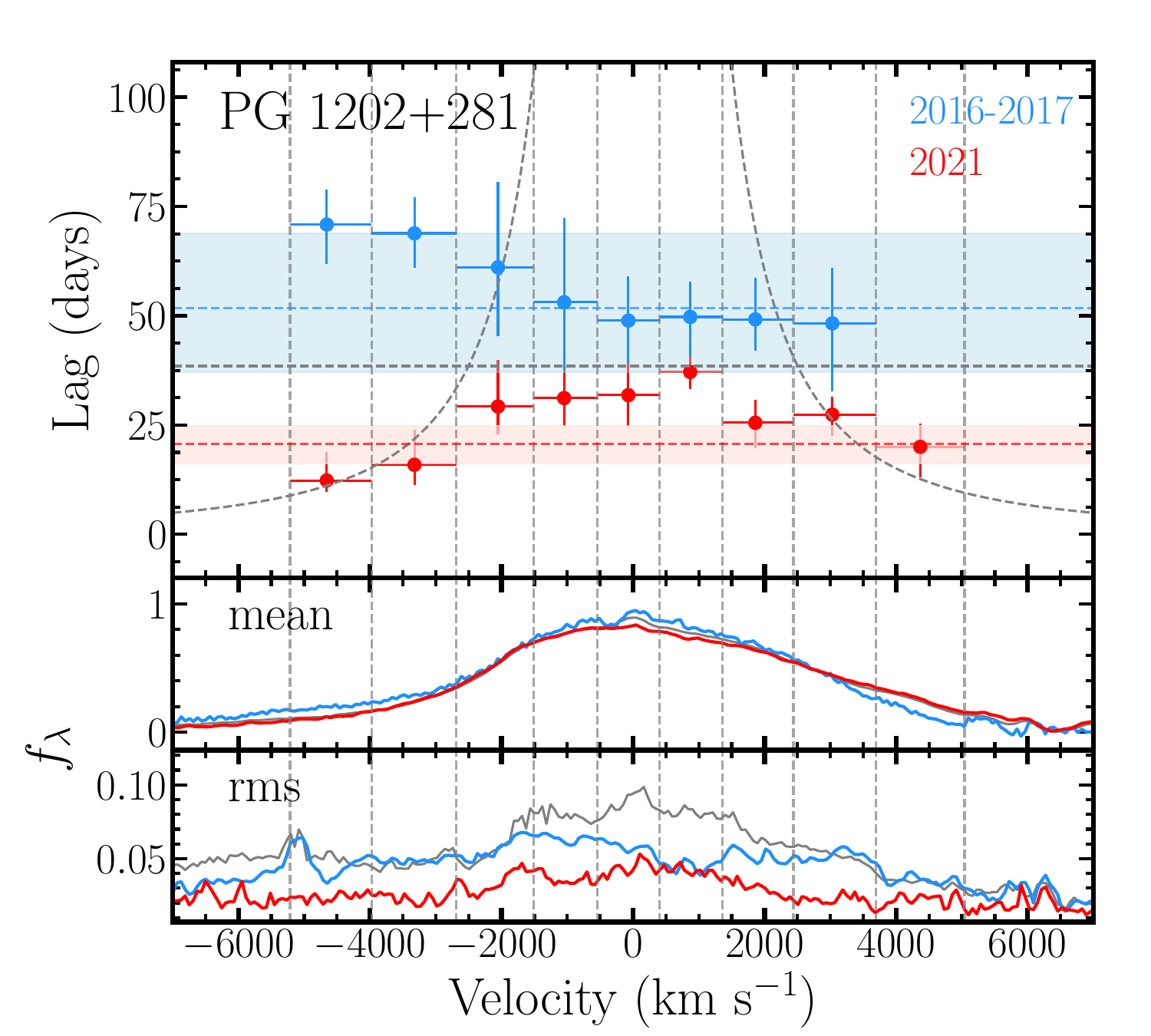}
                \includegraphics[width=0.45\textwidth]{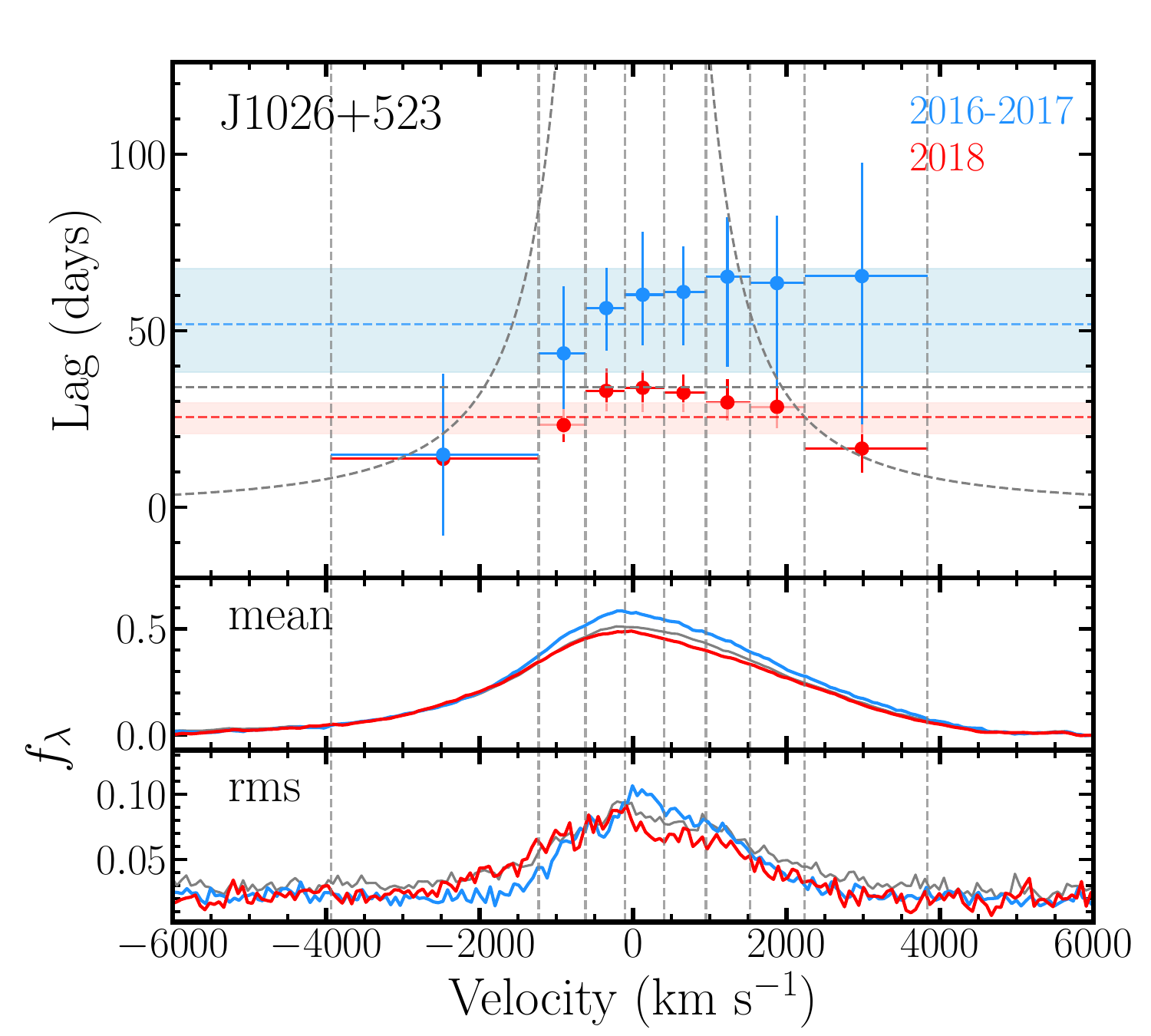}

   \caption{Comparison \hbeta\ velocity-resolved lags based on the observation in different years for two objects (left: PG 1202+281; right: J1026+523).  The corresponding mean and rms spectra in the unit of 10$^{-15}$ erg s$^{-1}$ cm$^{-2}$ ${\rm \AA}^{-1}$  are presented in the middle and bottom panels, respectively. The symbols are similar as those Figure \ref{fig:VRL-1}. The only difference is that the result in different year(s) are plotted by different colors as indicated at the upper-right corner.  The grey horizontal line represents the integrated \hbeta\ lag based on 6 yr observation.}
    \label{fig:VRL-Multiple-year}
\end{figure*}

For two AGNs (PG 1202+281 and J1026+523), we were able to measure the velocity-resolved lags in different years (see Figure \ref{fig:VRL-Multiple-year}), allowing us to study the evolution of BLR kinematics.

For PG 1202+281, we successfully measured velocity-resolved lags based on 2016--2017 and 2021 observations.  It displayed a dominated inflow-like signature in 2016--2017, while exhibiting a symmetric structure  in 2021, consistent with disk-like rotation.  The rms spectra also looked different, where the 2016--2017 rms spectra suggested more variation in the line wing while the 2021 rms spectrum indicated that the main variation occurred in the line center.  

The overall  \hbeta\ lag in the rest frame decreased from 52$^{+14}_{-17}$ days in 2016--2017 to 20$^{+5}_{-4}$ days in 2021 (a decreasing of 62\%), as the average continuum luminosity at 5100\AA\ ($L_{5100}$) decreased $\sim$20\%.  As a result, the slope of PG 1202+281's $R_{\rm BLR}$--$L_{5100}$ relation is significantly steeper than the expected value of 0.5 {as well as the observed slope of the ensemble $R_{\rm BLR}$--$L_{5100}$ relation \citep[e.g.,][]{Woo24,Wang24}.  We also notice that in 2021 there was a dip followed by a drastic flux increase during the last two months (see Paper \RNum{3}).  
The $\sigma_{\rm rms}$ was not well constrained in 2021 for this object. Using the mean FWHM instead, the estimated BH mass in 2021 is only 50\%  of than in 2016--2017. Given that the reported BH mass has a 25\% uncertainty (Paper \RNum{3}), there is a slight inconsistency in the BH mass between the two observations. If confirmed at higher significance, it would  suggest a change in the virial coefficient.

PG 1202+281 was also monitored by MAHA mainly from 2017 to 2020 \citep{Bao22}. They successfully measured  velocity-resolved lags in 2017, 2019, and 2020. Their results in 2017 showed a strong inflow-like signature, which is consistent with our findings based on 2016--2017 observations. In 2019, the inflow-like signature became less prominent with the longest lag shifted to zero velocity, although the red wing still appeared to have shorter lags than the blue wing. In 2020, the velocity-resolved lags were flat,  with only one central velocity bin showing a significantly longer lag. While our data in these two years could not well constrain the velocity-resolved lags, we find consistent trends between these two campaigns, confirming that the observed changes are genuine and not due to artifacts or systematics. 

By combining the data from the two campaigns, we obtain nearly yearly velocity-resolved lag measurements for this source from 2016/2017 to 2021. These results reveal significant evolution in velocity-resolved structures, transitioning from inflow-like to symmetric over $4\sim 5$ years in the rest frame. 

J1026+523  is another AGN that shows different velocity-resolved structures within our monitoring baseline. 
Its velocity-resolved lags showed a significant outflow-like structure in 2016--2017, while in 2018 it was more symmetric and  consistent with a disk-like rotation interpretation.   We should note that the 2016--2017 results have larger lag uncertainties compared to the results from 2018, especially in the red velocity bins. Therefore, the difference in lags in the red velocity bins is not significant considering the lag uncertainties. Nevertheless, there is some evidence supporting that the transition is real. For instance,  the overall results based on 6-year observations (Figure \ref{fig:VRL-1}) also show a red-lagging-blue structure but with smaller lag uncertainties.  In addition, the  2016--2017 result shows a gradual and  consistent trend from the line center to the red wing. Taking these points into account, we favor the scenario of an real transition. 

We construct the mean and rms spectra based on the observation from 2016--2017 and 2018, respectively.  Compared to 2018, the mean spectrum in 2016--2017 shows larger flux from the line center to the red wing. Similarly, the 2016--2017 rms spectrum also exhibits a red excess, along with a blue flux deficit.  On the contrary,  the 2018 rms spectrum is slightly more symmetric and broader compared to the 2016--2017 rms spectrum.  In addition,  the integrated \hbeta\ lags also drops from 52$^{+14}_{-16}$ days in 2016--2017 to 26$^{+5}_{-4}$ days in 2018 in the rest frame (51\% decrease),  as $L_{5100}$ decreases $\sim$10\%. Again, the slope of J1026+523's $R_{\rm BLR}$--$L_{5100}$ relation is significantly steeper than 0.5. Additionally, the $\sigma_{\rm rms}$ increases by a factor of 1.24, from 1013$\pm60$ km s$^{-1}$ in 2016--2017 to 1252$\pm48$ km s$^{-1}$ in 2018.  The derived BH mass is approximately consistent between 2016--2017 and 2018 considering that the BH mass has a 25\% uncertainty.

\section{Discussion} \label{sec:discussion}

\subsection{Demographics of BLR Kinematics}

In this section, we investigate the demographics of the BLR kinematics and its connection with AGN properties based on velocity resolved lag measurements. Combining the velocity-resolved lags from SAMP and literature \citep{Kollatschny03,Bentz09b,Denney09b,Denney10,Barth11,Doroshenko12,Grier13,Du16,Du18,Du18b,Lu16,Lu19, Lu21,DeRosa18,Zhang19, Hu20b, Hu20a, Brotherton20,Bentz21,Bentz23b,Bentz23a, Feng21b,Feng21a, Oknyansky21, U22,Bao22,Li-S22,Chen23,Zastrocky24}, we obtained a sample of 94 unique AGNs.  We assign equal weight to each AGN, and for AGNs with multiple measurements, we calculate the percentage of each type and incorporate it into the total count. 

Figure \ref{fig:VRL-pieplot} illustrates the demographics of the BLR kinematics based on the structure of velocity-resolved lags,  where  40\%, 20\%, and 11\% of our combined sample show symmetric, inflow-like and outflow-like structures, respectively. In the other 29\% cases, the velocity-resolved lags are ambiguous. 
If symmetric velocity-resolved lags can be interpreted as indicating disk-like rotation, this suggests that disk-like rotation is the most common BLR kinematics. The frequency of inflow is also notable, which may reflect the presence of ongoing accretion. The lower occurrence of outflow suggests that the outflow generally plays a less significant role in BLR kinematics. These results are qualitatively consistent with the findings from velocity-delayed maps and dynamical modeling studies \citep[e.g.,][]{Bentz09b, Pancoast14b, Grier17b,Xiao18a,Xiao18b,Williams18, Horne20,Villafana22,Villafana23} as well as the studies of BLR breathing effects based on \hbeta\ \citep[e.g.,][]{Peterson99,Peterson00,Park12,Barth15,Wang20}. 
Note that the literature sample mostly consists of low-to-moderate-luminosity AGNs. The SAMP velocity-resolved RM sample,  with the luminosity range of $L_{5100}\sim$[$10^{44.1}$, $10^{45.1}$] erg s$^{-1}$,  significantly supplements the previous velocity-resolved RM sample in high-luminosity regime.

\begin{figure}[htbp]
    \centering
    \includegraphics[width=0.43\textwidth]{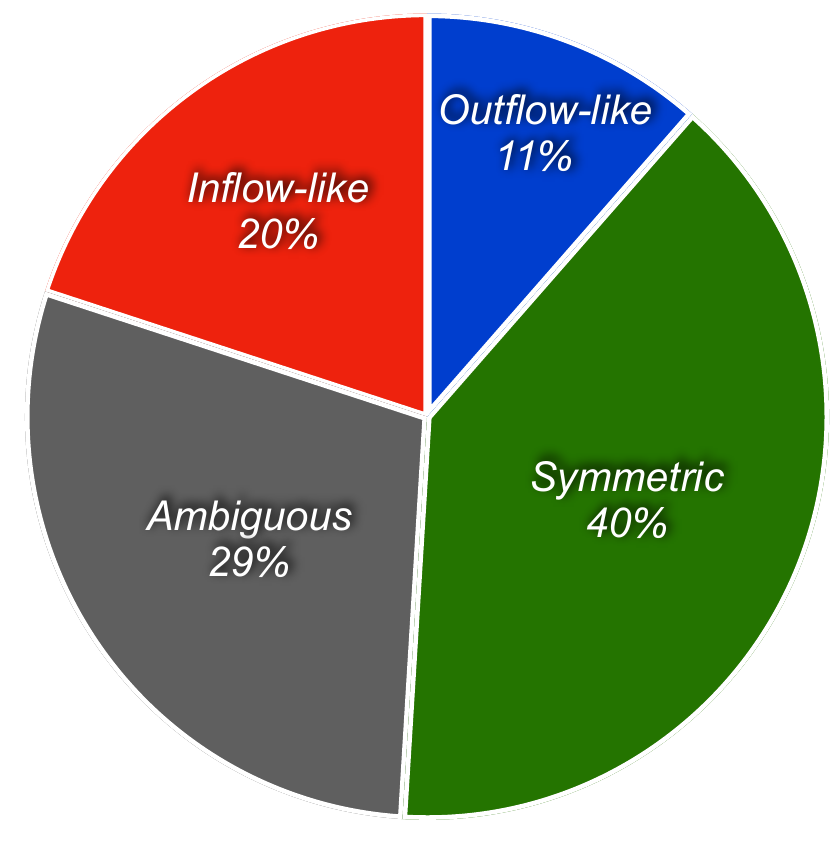}
              
   \caption{Demographics of the BLR kinematics based on the structure of velocity-resolved lags of 94 AGNs. Symmetric, inflow-like, outflow-like, and ambiguous velocity-resolved lags are presented in green, red, blue, and grey, respectively, along with the fraction of each category.}
    \label{fig:VRL-pieplot}
\end{figure}

In RM-based BH mass estimations, it is generally assumed that radiation pressure is negligible and the BH gravitational force  dominates. However, BLRs with outflow kinematics may be influenced by radiation pressure. A non-negligible radiation pressure can introduce a bias in the mass estimates, although this bias could  be already accounted for in the uncertainty of the virial factor. Our results show that the proportion of outflow kinematics in AGN BLRs is relatively small, indicating that the bias introduced by radiation pressure may not have a strong impact on the BH estimation, if radiation pressure drives outflows.

To further investigate the  trends between the inferred BLR kinematics and AGN properties, we first plot the velocity-resolved RM sample on the $R_{\rm BLR}$--$L_{5100}$ plane (Figure \ref{fig:VRL-Two-plane}). 
Only the sample with available host-corrected $L_{5100}$ is included. We find that symmetric and outflow-like velocity-resolved lags show no significant dependency on luminosity. Neither do they show a clear correlation with the offset from the $R_{\rm BLR}$--$L_{5100}$ relation. In comparison, there seems a potential aggregation of AGNs with inflow-like velocity-resolved lags around $L_{5100}\sim10^{44}$ erg s$^{-1}$. Note that some of these AGNs also exhibit significant deviations from the R--L relation.} Further investigation  with a larger sample is required. 
In addition, we investigate the connection between BLR kinematics and Eddington ratios ($\lambda_{\rm Edd}$), where the bolometric luminosity is calculated from $L_{5100}$ using a single conversion factor of 9.26 \citep[e.g.,][]{Shen11}.  We do not observe any clear trends between BLR kinematics with $\lambda_{\rm Edd}$ in the $L_{5100}$-$\lambda_{\rm Edd}$ plane. 

We also  investigate the trends on the FWHM--$R_{\rm Fe}$ plane in Figure \ref{fig:Third-plane}. The $R_{\rm Fe}$ is defined as the flux ratio between iron  (from 4434 to 4684 \AA) and broad \hbeta. The anti-correlation between \hbeta\ FWHM and $R_{\rm Fe}$ is  known as the AGN Eigenvector 1 sequence \citep{Boroson92}. The detailed physical process underlying this sequence is still subject to some level of debate \citep{Panda19}, but the primary drivers are thought to be $\lambda_{\rm Edd}$ and orientation \citep[e.g.,][]{Boroson92,Sulentic00,Shen14,Martinez19}.

We divide our sample into small and large $R_{\rm Fe}$ values based on $R_{\rm Fe}=0.8$. We find that sources with higher $R_{\rm Fe}$ might show a larger occurrence rate of outflow-like velocity-resolved lags, but the sample size of objects with large $R_{\rm Fe}$ is too small to draw any conclusions. In addition, we divide our sample based on FWHM  thresholds of 2000 km s$^{-1}$ and 4000 km s$^{-1}$. The former distinguishes broad-line and narrow-line Seyfert 1 galaxies  \citep{Osterbrock85,Goodrich89} while the latter separates the empirical AGN Population A and B \citep{Sulentic00}. We find no clear trend between kinematics and FWHM, though outflows appear less frequent in AGNs with large FWHM, suggesting a potential preference for lower inclination angles.

\begin{figure}[htbp]
    \centering
    \includegraphics[width=0.49\textwidth]{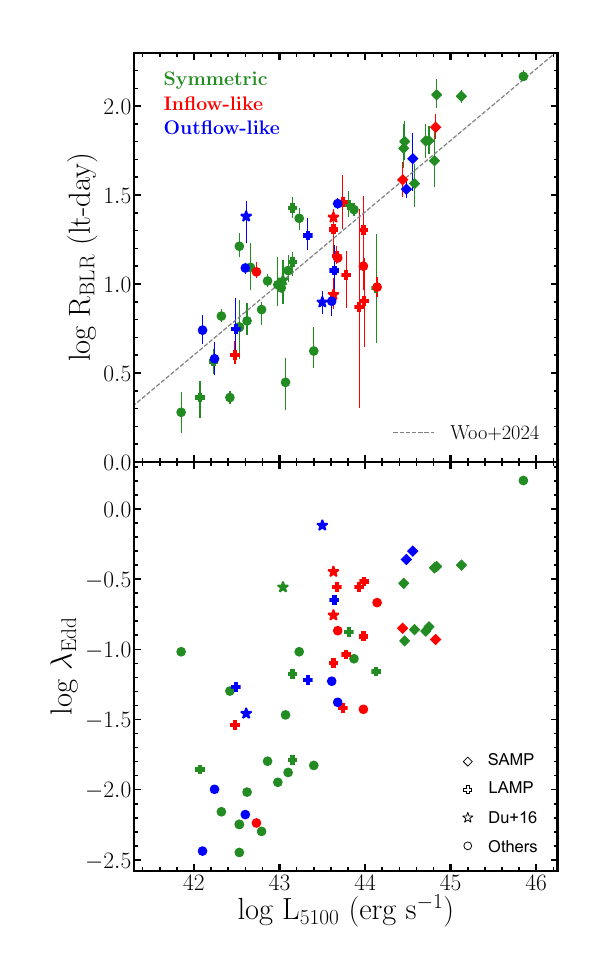}

   \caption{Trends of BLR kinematics on the $R_{\rm BLR}$--$L_{5100}$ plane (upper) and the $\lambda_{\rm Edd}$--$L_{5100}$ plane (lower). Green, red, and blue colors represent symmetric, inflow-like, and outflow-like velocity-resolved lags, respectively. We used a different symbol for each adopted sample (bottom-right corner in the lower panel). In the upper panel, we adopt the $R_{\rm BLR}$--$L_{5100}$ relation from \citet{Woo24} as the fiducial relation. }
    \label{fig:VRL-Two-plane}
\end{figure}

Note that the discussion in this section is based on the qualitative interpretation of the velocity-resolved lags. The velocity-resolved lags represent the averaged delays at certain line-of-sight velocities, and thus cannot fully constrain the BLR structure and kinematics. BLR kinematics are often a combination of both disk-like rotation and inflow/outflow components \citep[e.g.,][]{Villafana22}, so a single description may be overly simplistic. The visual inspection further introduces a substantial selection bias.  
All of these factors suggest the requirement for more detailed  quantitative approaches, such as dynamical modeling, to provide better understanding of BLR structure and kinematics.

\begin{figure}[htbp]
    \centering
    \includegraphics[width=0.49\textwidth]{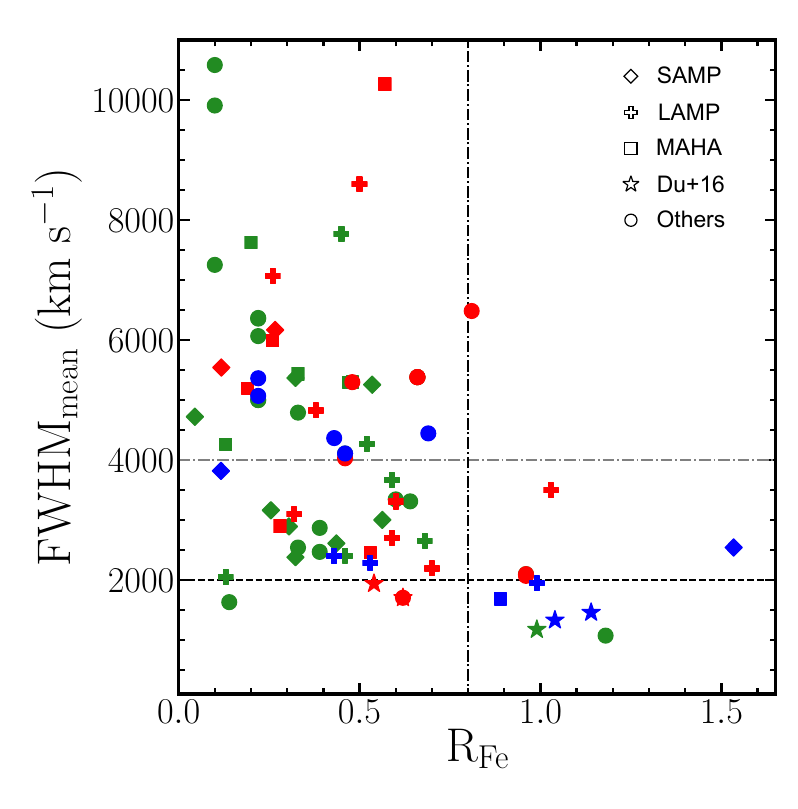}

   \caption{ Trends of BLR kinematics on the H$\beta$ FWHM--$R_{\rm Fe}$ plane. Different colors refer to different kinematics types and different symbols denote different samples as in Figure \ref{fig:VRL-Two-plane}. 
   The vertical dotted dashed line corresponds to $R_{\rm Fe}=0.8$. The horizontal dashed and dotted-dashed line represent FWHM$ = 2000$ and 4000 km s$^{-1}$, respectively.} 
    \label{fig:Third-plane}
\end{figure}

\subsection{BLR Kinematics Evolution}

In \S \ref{sec:changes_kinematics},  we discover that two AGNs in our sample exhibit different velocity-resolved structures over the 6 yr baseline. Similar evolution has also been reported in the literature \citep[e.g.,][]{DeRosa18, Bao22, Chen23,Zastrocky24,Feng24}, which may suggest that the changes in BLR kinematics on relatively short timescales are not rare.

A plausible hypothesis is that the BLR is composed of multiple components. For instance,  an inflow component was present in the BLR of PG 1202+281 from 2016 to 2019. As this inflow component moved inwards, its response progressively diminished. In 2021, only the disk-like rotational component remained to respond to the  variation in continuum flux.
Similarly, J1026+523's BLR may consist of a Keplerian disk and an outflow component.  During 2016--2017, the outflow component dominated the response of \hbeta, which, however, had faded or been expelled by 2018, leaving only the inner disk-like component to respond.  
A related discussion is that any global change of BLR should occur at the dynamical timescale \citep{Peterson93,Lu16}:
\begin{equation}
    t_{\rm dyn,BLR} = \frac{c\tau_{\rm H\beta}}{V_{\rm FWHM}} = 3.36 \frac{\tau_{20} }{V_{5000}} {\rm yrs}
\end{equation}
where $V_{5000}$ is the FWHM of \hbeta\ normalized by 5000 km s$^{-1}$ and $\tau_{20}$ is the rest-frame lag normalized by 20 days.  In some low-luminosity AGNs, e.g., NGC 3227, the dynamical time scale is short ($\sim$ 1 yr) and smaller than the observed variation.   For high-luminosity AGNs like J1026+523 and PG 1202+281, the $t_{\rm dyn,BLR}$ is longer, e.g., $7.2$ and 5.6 yrs, respectively. The observed transition in rest frame is comparable or shorter than the dynamical timescale.

It should be noted that the changes in velocity-resolved lags do not uniquely prove the changes in BLR kinematics. They could be attributed to other factors, such as varying obscuration from the outflowing dusty clumps within the BLR \citep{Gaskell18}, or variations in the distribution of BLR gas density and ionization states. For instance, if spiral arms exist inside BLRs, the rotation of these spiral arms can lead to significant changes in the line profiles as well as the velocity-resolved lags within relatively short time \citep{Wang-J22,Du23}, i.e., $<10$ years. 
More detailed  modeling considering both the gas density and the emissivity is necessary for understanding the changes in velocity-resolved structures \citep{Williams22, Lizvette24}.

\section{Conclusion} \label{sec:conclusion}

We present the \hbeta\ velocity-resolved lag measurements for 20 AGNs from SAMP.  Among these 20 sources, we identify 12 AGNs with significant resolved and unambiguous structures based on visual inspection and quantitative methods. We find that eight objects exhibit symmetric/center-lagging-wing structures, two objects show inflow-like characteristics, and two objects demonstrate outflow-like signatures. For the remaining eight objects with ambiguous trends, we leave their kinematic interpretation as indeterminate.

For two AGNs, PG 1202+281 and J1026+523, we successfully obtained velocity-resolved lags in different years. The velocity-resolved lags of PG 1202+281 exhibited strong inflow-like signatures in 2016--2017, which changed to symmetric by 2021. Our result aligns with the trends found by MAHA's observations during 2017 to 2020. Combining the results from two campaigns together, the nearly yearly velocity-resolved lag measurements of PG 1202+281 suggests a gradual transition from inflow  to disk-like rotation  from 2016 to 2021. J1026+523 showed outflow-like velocity-resolved lags in 2016--2017, which shifted to symmetric in 2018,  possibly due to a fading outflow component.  These results may indicate that BLR kinematics can change over relatively short timescales, even in these high-luminosity AGNs whose dynamical timescale is longer than the lower-luminosity AGNs.

Combining the velocity-resolved lags from SAMP and literature, we analyzed demographics of the inferred BLR kinematics and the connection with AGN properties using 94 unique AGNs. We found that symmetric velocity-resolved structure is the most commonly observed type,  comprising 40\% of cases. The frequency of inflow-like structure is also notable (20\%), indicating ongoing accretion in many AGNs, while the outflow-like signature is less common (11\%), suggesting a less significant role in BLR kinematics. In the other 29\% cases,  the velocity-resolved lags are ambiguous and kinematics are indeterminate. We observe no obvious trends between the inferred kinematics with  $L_{5100}$ or $\lambda_{\rm Edd}$, except for a potential aggregation of AGNs with inflow kinematics at $L_{5100}\sim10^{44}$ erg s$^{-1}$. Sources with higher $R_{\rm Fe}$ might show a larger occurrence rate of outflow kinematics, but the sample size of objects with large $R_{\rm Fe}$ is too small for any conclusions. Note that these results are based on qualitative interpretations of velocity-resolved lags. More detailed dynamical modeling is necessary for more accurate constraints of  BLR kinematics.

The SAMP sample, characterized by moderate to high luminosities, substantially supplements the previous velocity-resolved RM sample in the high-luminosity regime. Our investigation demonstrates the use of qualitative velocity-resolved lags to improve our comprehension of BLR kinematics. Future follow-up dynamical modeling (S. Wang et al., in preparation)  of this sample will allow for more detailed studies of their BLR structure and kinematics.

\section{Acknowledgment}

This work is supported by the National Research Foundation of Korea (NRF) grant funded by the Korean government (MEST) (No. 2019R1A6A1A10073437 and No. 2021R1A2C3008486).   The research at UCLA was supported by the NSF grant NSF-AST-1907208. The research at UC Irvine was supported by NSF grant AST-1907290. V.N.B. gratefully acknowledges assistance from NSF Research at Undergraduate Institutions (RUI) grant AST-1909297. Note that the findings and conclusions do not necessarily represent the views of the NSF. V.U acknowledges funding support from NSF Astronomy and Astrophysics Research Grant No. AST-2408820, NASA Astrophysics Data Analysis Program (ADAP) grant No. 80NSSC23K0750, and STScI grant Nos. HST-AR-17063.005-A, HST-GO-17285.001-A, and JWST-GO-01717.001-A.  We thank the useful discussion with D.W. Bao.

This work utilizes the data from the ZTF. ZTF is supported by the National Science Foundation under Grant No. AST-2034437 and a collaboration including Caltech, IPAC, the Weizmann Institute for Science, the Oskar Klein Center at Stockholm University, the University of Maryland, Deutsches Elektronen-Synchrotron and Humboldt University, the TANGO Consortium of Taiwan, the University of Wisconsin at Milwaukee, Trinity College Dublin, Lawrence Livermore National Laboratories, and IN2P3, France. Operations are conducted by COO, IPAC, and UW. This research has made use of the NASA/IPAC Infrared Science Archive, which is funded by the National Aeronautics and Space Administration and operated by the California Institute of Technology.

\bibliography{ref}

\appendix

\section{Comments on individual object} \label{sec:comments}

\textit{Mrk 1501}:  The velocity-resolved lags are ambiguous given the large lag uncertainties. The two rightmost bins have negative lags and are removed, along with the leftmost bin due to large uncertainty. This object was monitored in 2010 by \citet{Grier13}, which revealed an inflow-like signature. It was also monitored by MAHA from 2018 to 2021 \citep{Bao22},  which showed a symmetric structure indicative of a transition from inflow to disk-like rotation in about 10 yr. Unfortunately, our data quality is not sufficient to confirm this transition.  

\textit{PG 0052+251}:   The velocity-resolved lags display a slight red-lagging-blue structure, though primary feature is  a clear center-lagging-wing pattern.  It may indicate the presence of a combination of disk-like rotation and outflow in PG 0052+251. 

\textit{J0101$+$422}: The results in Table \ref{tab:BLR-kinematics} are measured using the 2018 light curves. While its $\chi^2<2$, we found a consistent trend among different velocity bins based on visual inspection, which shows an  inflow-like structure.   Interestingly,  the rms spectrum shows a blueshift peak, while the mean spectrum is not blueshifted.  We also checked the results based on 6 yr light curves. We found a different result that it generally show a center-lagging-wing and red-lagging-blue structure. However, after shifting the continuum light curve based on measured lags to check the match between continuum and velocity-resolved \hbeta\ light curves, we found that the lags based on 6-yr light curves provide a less convincing match compared to those based on 2018 observation only. Consequently, we reported the velocity-resolved lags in 2018 in Table \ref{tab:BLR-kinematics} .

\textit{J0140$+$234}: It has a ${\Delta \tau}_{\rm corr} / \overline{\tau}>0$ suggesting significantly resolved lags. It shows a center-lagging-wing structure,  which indicates the presence of disk-like rotation. The rightmost bin has a relatively noisy light curve but cannot be excluded based on our criteria.  Taking this into account, the presence of inflow in this object is also suggested.

\textit{J0939$+$375}: The lags are measured using the 2018 light curves, following Paper \RNum{3}. The two leftmost bins and the far-right bin are excluded because their lags are consistent with zero. The lags in the remaining velocity bins are not resolved, leaving the BLR kinematics indeterminate. 

\textit{PG 0947$+$396}: The velocity-resolved lags generally show a center-lagging-wing structure, indicative of disk-like rotation.  It also displays a clear red asymmetry,  i.e.,  the lags at blue velocity bins are generally larger than those in corresponding red velocity bins. This signature may also indicate the presence of inflow.  These findings and interpretations are consistent with those by \citet{Bao22}, where the authors monitored PG 0947$+$396 from 2018 to 2021.  

\textit{J1026+523}: J1026+523 has been discussed in section \ref{sec:changes_kinematics}. 

\textit{J1120$+$423}: The velocity-resolved lags display a complicated  double-peak structure.  We leave the BLR kinematics of J1120$+$423 as indeterminate.

\textit{PG 1121+422}: While most of the lags are within 1$\sigma$ uncertainty of the integrated \hbeta\ lag, the lag uncertainty of each bin is small. Therefore,  it has a ${\Delta \tau}_{\rm corr} / \overline{\tau}>0$.  The center most bin displays the longest lag, which may suggest the presence of disk-like rotation. However, inflow kinematics could also present because the rightmost bin shows a much smaller lag compared the leftmost bin. 

\textit{PG 1202+281}: PG 1202+281 has been discussed in section \ref{sec:changes_kinematics}.  

\textit{J1217$+$333}: The lag at the rightmost bin is negative and  is therefore excluded. The lags in the two leftmost bins are also excluded because of low correlation strength ($r_{\rm max}<0.5$). Because there are insufficient valid bins on the blue side,  we set the interpretation as indeterminate. 

\textit{V\RNum{3}~Zw~218}:  The velocity-resolved lags are derived based on 2019--2021 observations.  It has a ${\Delta \tau}_{\rm corr} / \overline{\tau}=0.26$, suggesting a clearly resolved structure. It shows a clear center-lagging-wing structure, which may indicate the presence of disk-like rotation in its BLR. 

\textit{PG 1322+659}: The leftmost bin is excluded because the lag is not converged. Additionally, two bins with low correlation strength ($r_{\rm max}<0.5$) are removed. While it shows ${\Delta \tau}_{\rm corr} / \overline{\tau}=0$,  our visual inspection reveals a center-lagging-wing  structure, which suggests the presence of disk-like rotation. This result is consistent with the small $\chi^2_{\rm asy}$ value.  Note that the peak of both the velocity-resolved lags and rms spectra is redshifted for this object. Thus, we use the rms peak velocity as the mirror to match blue and red velocity bins in calculating $\chi^2_{\rm asy}$.

\textit{PG 1427+480}: The velocity-resolved lags are measured based on 2018--2021 observations due to cadence consideration. While it shows a slight blue-lagging-red pattern, the difference is not significant compared to the uncertainty. In particular,  the central three bins do not show a consistent blue-lagging-red pattern. We consider the structure as ambiguous and leave the interpretation as indeterminate.  

\textit{PG 1440+357}: The leftmost bin has a low correlation coefficient of $r_{\rm max}<0.5$ and a lag of $\sim$300 days, which is excluded from the analysis. Visual inspection suggests there is a red-lagging-blue structure, which indicates that the BLR kinematics are likely outflow.  

\textit{J1456+380}: The leftmost and rightmost bin are removed because of low correlation strength. While there could be two peaks at $V\sim-2500$ and $V\sim1200$ km s$^{-1}$ in the velocity-resolved lags, the difference among the rest velocity bins is not significant. We consider the structure as flat and leave the interpretation as indeterminate. 

\textit{J1540+355}: While it has a ${\Delta \tau}_{\rm corr} / \overline{\tau}=0$, the velocity-resolved shows a clear center-lagging-wing structure, which can be interpreted as disk-like rotation.

\textit{PG 1612+261}: The velocity-resolved lags are measured based on the 2021 observation which contains most observations for this object. The velocity-resolved lag shows a clear center-lagging-wing pattern, which indicates the presence of disk-like rotation. We interpret the BLR kinematics as disk-like rotation. 

\textit{J1619+501}:  The lag in the leftmost bin is not converged and is excluded in the analysis.  We find that only one central velocity bin shows a significantly larger lag than the others. We consider the structure as ambiguous and leave the interpretation as indeterminate.

\textit{PG 2349$-$014}: It shows a ${\Delta \tau}_{\rm corr} / \overline{\tau}=0$, and an ambiguous structure. We leave the interpretation as indeterminate.   

\end{document}